\documentclass{aa}  

\usepackage[dvipsnames]{xcolor} 
\usepackage{graphicx}
\usepackage{txfonts}
\usepackage[colorlinks=true, linkcolor=blue, urlcolor=blue, citecolor=blue]{hyperref}
\usepackage{orcidlink}

\defcitealias{Flaischlen2021}{F21}

\begin{document}

   \title{Monitoring accretion rate variability in the Orion Nebula 
   Cluster with the Wendelstein Wide Field Imager\thanks{The data underlying the Tables \ref{tab:tabphot1}, \ref{tab:tabphot2}, and \ref{tab:tabparams} and the reduced WWFI data are only available at the CDS
via anonymous ftp to \url{cdsarc.u-strasbg.fr} (\url{130.79.128.5}) or via
\url{http://cdsarc.u-strasbg.fr/viz-bin/cat/J/A+A/xxx/Axxx}}}
   \titlerunning{WWFI accretion monitoring}
   \author{S. Flaischlen\inst{1}
                        \and
          T. Preibisch\orcidlink{0000-0003-3130-7796}\inst{1}
                        \and
												M. Kluge\orcidlink{0000-0002-9618-2552}\inst{1}
												\and
                  C.F. Manara\orcidlink{0000-0003-3562-262}\inst{2}
                        \and
                  B. Ercolano\inst{1}}
   \institute{Universitäts-Sternwarte, Fakultät für Physik, Ludwig-Maximilians-Universität München, 
                        Scheinerstr. 1, 81679 München, Germany
         \and
             European Southern Observatory, 
                         Karl-Schwarzschild-Str. 2, 85748 Garching bei München, Germany
             }

   \date{Received November 9, 2021; accepted July 25, 2022}

 
  \abstract
   {The understanding of the accretion process has a central role in the understanding of star and planet formation. }
   {We aim to test how accretion variability influences previous correlation analyses of the relation between X-ray activity and accretion rates, which is
	  important for understanding the evolution of circumstellar disks and disk photoevaporation.
      }
   {We monitored accreting stars in the Orion Nebula Cluster from November 24, 2014, until February 17, 2019, for 42 epochs with the \textit{Wendelstein Wide Field Imager} in the Sloan Digital Sky Survey $u^\prime g^\prime r^\prime$ filters on the 2 m Fraunhofer Telescope on Mount Wendelstein. Mass accretion rates were determined from the measured ultraviolet excess. The influence of the mass accretion rate variability on the relation between X-ray luminosities and mass accretion rates was analyzed statistically.
        }
   {
        We find a typical interquartile range of $\sim 0.3~\mathrm{dex}$ for the mass accretion rate variability on timescales from weeks to $\sim 2$ years. 
The variability has likely no significant influence on a correlation analysis of the X-ray luminosity and the mass accretion rate observed at different times when the sample size is large enough.
        }
   {The observed anticorrelation between the X-ray luminosity and the mass accretion rate predicted by models of photoevaporation-starved accretion is likely not due to a bias introduced by different observing times.
		}

   \maketitle


\section{Introduction}

The accretion rate is a crucial parameter for the understanding of accretion disks surrounding young stellar objects (YSOs) and planet formation \citep[e.g.,][]{Ercolano2017}. The interaction between the circumstellar disk and the central star is thought to be regulated by magnetic field lines channeling the flow of material along accretion columns from the inner edge of the disk to the stellar surface \citep{Hartmann2016}. Accretion shocks produced when the infalling material hits the photosphere with nearly free-fall velocity produce a variety of observational features like optical and ultraviolet (UV) excess emission, spectral veiling, and strong optical emission lines. From these features, the accretion luminosity, $L_\mathrm{acc}$, can be determined from spectroscopy \citep[e.g.,][]{Alcala2017} or photometry \citep[e.g.,][]{Manara2012}.

The accretion rate tends to decrease over time \citep[e.g.,][]{Hartmann1998} as the disk material gets accreted or dispersed by disk winds, driven by irradiation from the star \citep{Alexander2014}, although several recent results show that accretion rates remain high in a number of older objects 
\citep[e.g.,][]{Ingleby2014, Rugel2018, Venuti2019, Manara2020}. Simulations suggest that X-ray emission is very efficient in photoevaporating the disk material \citep[e.g.,][]{Ercolano2008b, Ercolano2008a, Picogna2019, Ercolano2021} and may decrease the accretion rate \citep[e.g.,][]{Drake2009}.

Our statistical analysis of $332$ accreting young stars in the Orion Nebula Cluster (ONC) \citep[][hereafter \citetalias{Flaischlen2021}]{Flaischlen2021} suggests a weak anticorrelation between residual X-ray luminosities and accretion rates, as expected in the context of X-ray driven photoevaporation and in accordance with previous studies \citep{Telleschi2007, Drake2009}.

However, a possible caveat in the interpretation of the data stems from the temporal variability of YSOs over a wide range of wavelengths. The X-ray luminosity and the accretion rate are both subject to temporal variability \citep[e.g.,][]{Wolk2004, Venuti2014, Venuti2021}. Since the X-ray data were obtained in January 2003 \citep{Preibisch2005} and the Hubble Space Telescope (HST) data, on which the accretion rates are based, between October 2004 and April 2005 \citep{Robberto2013}, possible effects of the variability on timescales of a few years on the correlation analysis cannot be ruled out.

The accretion rate variability manifests itself mainly in the optical and near-UV bands \citep[e.g.,][]{Venuti2014, Venuti2015, Robinson2019, Schneider2020}. The observed brightness variations stem from a multitude of mechanisms, including accretion bursts \citep[e.g.,][]{Venuti2014}, inner disk warps \citep[e.g.,][]{Frasca2020}, accretion columns creating hot spots \citep[e.g.,][]{Cody2014} moving in and out of view while the YSO rotates \citep[e.g.,][]{Cody2018, Schneider2020} and phases of stable or unstable accretion regimes \citep[e.g.,][]{Sousa2016}.

In recent years, an abundance of effort has been made regarding variability analyses of YSOs. They were monitored with ground-based facilities \citep[e.g.,][]{Costigan2014, Venuti2014, Venuti2021} and space-based instruments such as the Convection, Rotation and Planetary Transits (\textit{CoRoT}) satellite \citep[e.g.,][]{Venuti2017}, the \textit{Kepler} space telescope in its \textit{K2} mission \citep[e.g.,][]{Pouilly2020, Rebull2020}, the \textit{Transiting Exoplanet Survey Satellite} (\textit{TESS}) \citep[e.g.,][]{Thanathibodee2020}, and the Microvariability \& Oscillations of STars (\textit{MOST}) space telescope \citep[e.g.,][]{Sousa2016, Siwak2018}.

The UV excess is a rather direct tracer of the accretion rate \citep[e.g.,][]{Schneider2020}. In this study, we are interested in characterizing the accretion rate variability through the UV excess of a particular sample of young stars in the ONC in order to analyze how this variability influences our statistical analysis of the relation between X-ray luminosities and accretion rates on timescales of weeks to a few years. 

To this aim, we performed a multiyear and multicolor photometric monitoring of the ONC with the wide-angle camera (the Wendelstein Wide Field Imager) at the 2~m Fraunhofer Telescope located on Mount Wendelstein in Germany \citep{Kosyra2014}. The monitoring provided $u^\prime$, $g^\prime$, and $r^\prime$ band photometry for several sources over the course of $\sim 4$ years, from which accretion rates could be estimated for a time span of $\sim 2$ years. The data were used to analyze how the accretion rate variability may influence the correlation analysis between X-ray activity and accretion and to characterize the accretion rate variability.

In Sec. \ref{sec:secobs}, we describe the observation and data reduction steps of
our WWFI data. Sec. \ref{sec:method} illustrates our method for
deriving stellar parameters and accretion rates from the obtained
photometry. In Sec. \ref{sec:secresults}, we characterize the accretion rate variability
and analyze how it influences the correlation analysis between X-ray luminosities and accretion rates.


\section{Observation and data reduction} \label{sec:secobs}

\begin{figure}
\resizebox{\hsize}{!}
        {
                \includegraphics[width=\hsize]{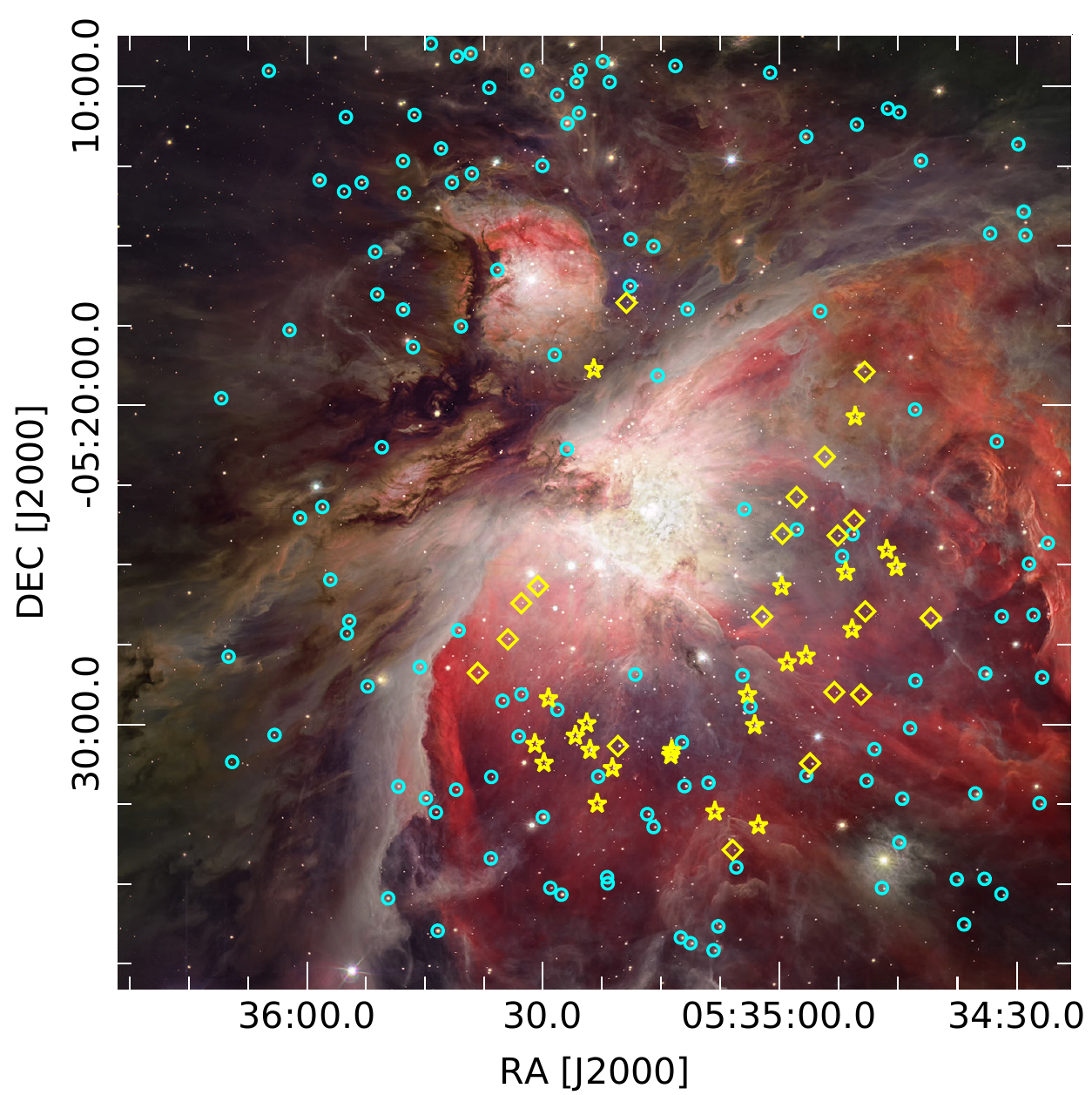}
        }
        \caption{WWFI $u^\prime g^\prime r^\prime$ color composition of the ONC (image credit: Wendelstein Observatory). 
				Overlaid are the stars monitored in all three bands. The yellow symbols indicate stars that have a match with the
				X-ray sources listed in the COUP catalog (J/ApJS/160/319/coup, \cite{Getman2005}). The star symbols further indicate sources that were
				monitored for at least five epochs and allowed an estimate for the mass accretion rate.} \label{fig:figorion}
\end{figure} 

\subsection{Target region}
The ONC is among the nearest regions of ongoing star formation \citep{Bally2008}, with a distance of $\sim 403~\mathrm{pc}$ \citep{Kuhn2019}. Therefore, it serves as a benchmark for star formation and has been studied in detail over the years. As such, stellar parameters for a large fraction of the young stellar population are available in the literature \citep{Manara2012}. Especially relevant for this work are the spectral type determinations from \cite{DaRio2012} and \cite{Hillenbrand2013}. Together with our photometric observation detailed below, the spectral types allow an estimate of the accretion rates.

\subsection{Data reduction and photometric calibration}

We observed the ONC in a multiyear and multicolor monitoring with the 2~m Fraunhofer Telescope at the Wendelstein Observatory located on Mount Wendelstein in Germany \citep{Kosyra2014}. The telescope hosts the \textit{Wendelstein Wide Field Imager} (WWFI), a wide field camera consisting of four $4096~\mathrm{px} \times 4109~\mathrm{px}$ charge-coupled device (CCD) detectors arranged in a $2 \times 2$ mosaic pattern. With $27.6\arcmin \times 28.9\arcmin$, the field of view covers the region of the X-ray data ($17\arcmin \times 17\arcmin$) used in our previous analysis of the relation between X-ray luminosities and accretion rates in \citetalias{Flaischlen2021}, which was based on data from the \textit{Chandra Orion Ultradeep Project} (COUP) \citep{Getman2005}.

For the monitoring, the $u^\prime$, $g^\prime$, and $r^\prime$ filters were used, which are approximately comparable with the photometric system of the \textit{Sloan Digital Sky Survey} \citep{Fukugita1996}. Since the central wavelength of the $u^\prime$ filter is close to the Balmer Jump, it is particularly sensitive for accretion effects. Figure \ref{fig:figorion} shows a WWFI composite image of the ONC taken with WWFI. The uncertainty of the $u^\prime$ photometry in the center region is large due to the very bright nebulosity.

The monitoring began on November 24, 2014, and comprises $42$ epochs until February 17, 2019. In order to fill up the gaps of the CCD mosaic, a $13$-step spiral dithering pattern of five exposures for each filter was utilized. With single exposure times of $30~\mathrm{s}$ ($u^\prime$), $20~\mathrm{s}$ ($g^\prime$) and $10~\mathrm{s}$ ($r^\prime$), sources with a brightness in the range of  $11.6 - 20.8~\mathrm{mag}$ in the $u^\prime$ band and $12.1 - 21.0~\mathrm{mag}$ ($11.6 - 20.8~\mathrm{mag}$) in the $g^\prime$ band ($r^\prime$ band) could be detected with a sufficient signal to noise ratio and without saturation. The mean seeing amounts to $1.60\arcsec \pm 0.29\arcsec$ for the $u^\prime$ band, $1.39\arcsec \pm 0.28\arcsec$ for the $g^\prime$ band, and $1.19\arcsec \pm 0.29\arcsec$ for the $r^\prime$ band. Due to weather conditions and technical maintenance, not all bands could be observed at each epoch. Table \ref{tab:tabepochs} in the appendix lists at which date how many sources were extracted for each filter.

\begin{figure}
\resizebox{\hsize}{!}
        {
                \includegraphics[width=\hsize]{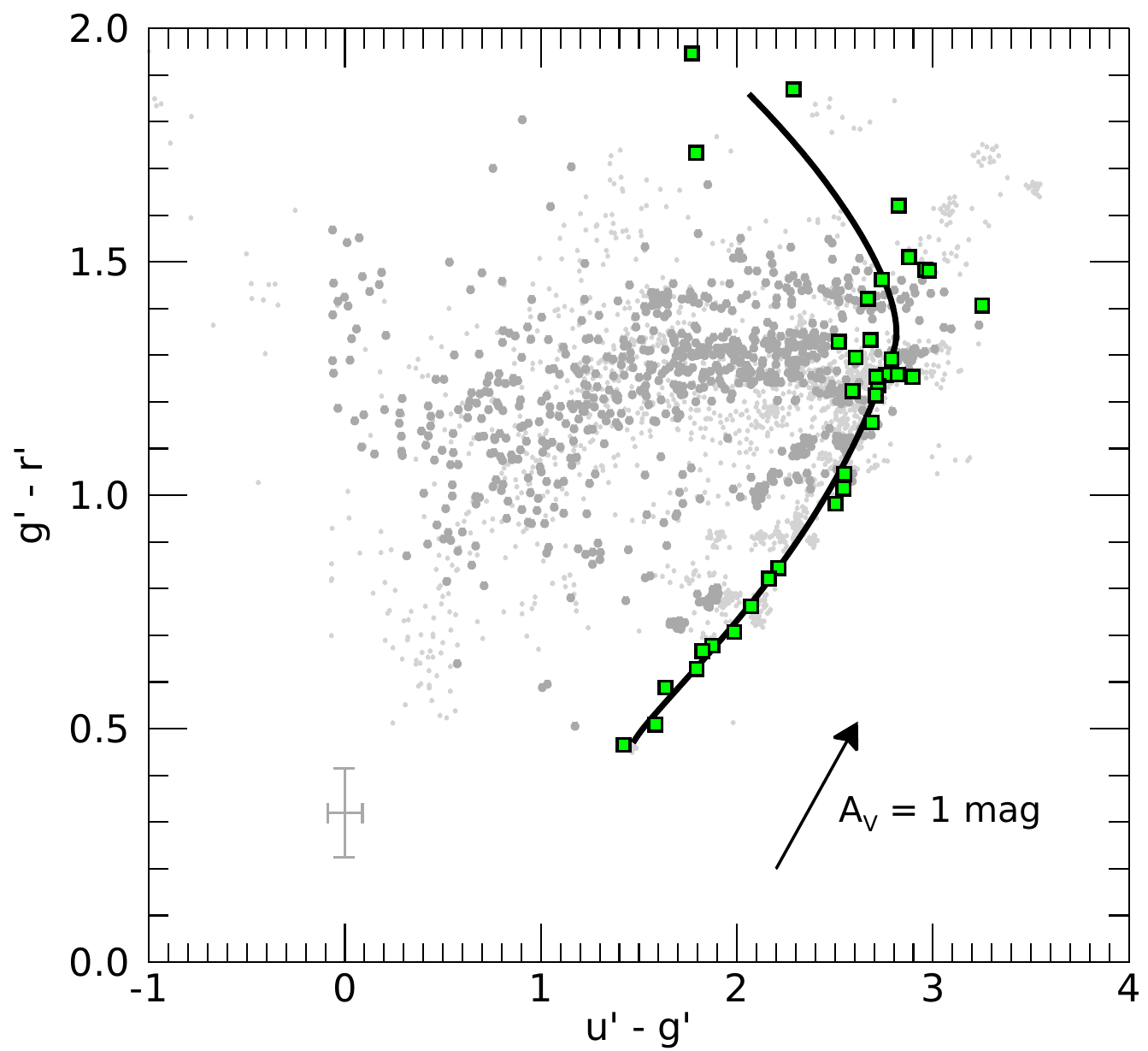}
        }
        \caption{$g^\prime - r^\prime$ vs. $u^\prime - g^\prime$ color-color diagram of YSOs monitored with WWFI as gray dots. The bigger dots indicate the objects displaying accretion according to \cite{Manara2012}. The gray cross shows the typical uncertainty of the photometry. The thick line is the sequence tracing the colors of the photosphere, obtained by synthetic photometry on spectra of diskless class III objects observed with X-Shooter \citep{Manara2013, Manara2017b}. The colors of these template spectra are shown as green squares. The arrow traces the reddening vector for $A_V = 1~\mathrm{mag}$, assuming the reddening law of \cite{Cardelli1989} and a galactic reddening parameter of $R_V = 3.1$.
} \label{fig:fig2cd}
\end{figure}

The raw data were reduced using the WWFI pipeline (see \cite{Kluge2020} for a more detailed description of the typical reduction steps). The data reduction includes bias subtraction, bad pixel, cosmic ray and satellite masking as well as flat fielding. The instrumental magnitudes were converted to AB magnitudes and the zero-points were calibrated by comparing the magnitudes with Pan-STARRS \citep{Flewelling2020} sources in the field of view for the $g^\prime$ and $r^\prime$ filters. In order to reduce the influence of variations due to different weather and instrumental conditions, the zero-points for each frame were further calibrated by aligning the magnitudes of non-variable sources with the corresponding magnitudes in a reference frame. Since there is no equivalent to the $u^\prime$ filter in the Pan-STARRS catalog, we calibrated the zero-point by fitting the color locus of non-accreting stars in the $g^\prime - r^\prime$ versus $u^\prime - g^\prime$ color-color diagram to the line describing the photospheric colors determined via synthetic photometry on empirical spectra of diskless class III objects using the $u^\prime$ passband from the WWFI setup. In Sec. \ref{sec:methphototrack}, we explain in more detail how we obtained this line. Figure \ref{fig:fig2cd} shows the color-color diagram for all targets and epochs.

Finally, the obtained magnitudes and photometric uncertainties for the three filters for each target and epoch were compiled into a single catalog, where we have limited ourselves to entries with photometry available simultaneously in all three bands and an uncertainty in the $u^\prime$ band magnitude $< 0.1~\mathrm{mag}$. The catalog contains $169$ stars with an average of $\sim 14.5$ $u^\prime$, $g^\prime$, and $r^\prime$ observations per star.

\begin{table*}
\caption{\label{tab:tabphot1}YSOs observed with WWFI and effective temperatures, extinctions and distances from the literature.}
\centering
\begin{tabular}{lccrrrl}
\hline\hline\\[-2.0ex]
Object\tablefootmark{(1)} & $\alpha_{2000}$  & $\delta_{2000}$ & \multicolumn{1}{c}{$T_\mathrm{eff}$} & 
\multicolumn{1}{c}{$A_V$\tablefootmark{(2)}} &
\multicolumn{1}{c}{Distance\tablefootmark{(3)}} & Notes \\
            &  $[$h:m:s$]$ & $[^{\circ}:\arcmin:\arcsec]$    & \multicolumn{1}{c}{$[$K$]$} & 
	\multicolumn{1}{c}{$[\mathrm{mag}]$} &		\multicolumn{1}{c}{$[\mathrm{pc}]$} &  \\[0.4ex]
\hline\\[-1.6ex]
\object{V397 Ori} & $05$:34:27.93 & $-05$:26:34.6 & $4205$ & $0.28$ & $400.5_{-4.5}^{+4.0}$ & 4 \\
\multicolumn{1}{c}{\vdots} & \multicolumn{1}{c}{\vdots} & \multicolumn{1}{c}{\vdots} & \multicolumn{1}{c}{\vdots} & \multicolumn{1}{c}{\vdots} & \multicolumn{1}{c}{\vdots} & \multicolumn{1}{l}{\vdots} \\[1.0ex]
\object{V1313 Ori} & $05$:34:43.41 & $-05$:30:07.0 & $3849$ & $0.97$ & $387.8_{-7.6}^{+7.0}$ & 5 \\[1.0ex]
\object{V1118 Ori} & $05$:34:44.74 & $-05$:33:42.1 & $3302$ & $2.06$ &  & 5, 6 \\[1.0ex]
\object{V1444 Ori} & $05$:34:45.19 & $-05$:25:04.1 & $4899$ & $0.96$ & $388.5_{-2.3}^{+2.0}$ & 5 \\[1.0ex]
\multicolumn{1}{c}{\ldots} & \multicolumn{1}{c}{\ldots} & \multicolumn{1}{c}{\ldots} & \multicolumn{1}{c}{\ldots} & \multicolumn{1}{c}{\ldots} & \multicolumn{1}{c}{\ldots} & \multicolumn{1}{l}{\ldots} \\[1.0ex] \hline
\end{tabular}
\tablefoot{
\tablefoottext{1}{Name of the YSO as listed in SIMBAD.}
\tablefoottext{2}{Extinctions from \cite{Manara2012}.}
\tablefoottext{3}{Distance calculated by inverting the Gaia EDR3 parallax.}
\tablefoottext{4}{Effective temperatures calculated with the temperature scales of \cite{Kenyon1995} (K type stars) and \cite{Luhman2003} (M type stars) with the spectral types from \cite{Hillenbrand2013}.}
\tablefoottext{5}{Effective temperatures taken from \cite{DaRio2012}.}
\tablefoottext{6}{Distance set to the mean distance of the ONC ($403~\mathrm{pc}$) determined by \cite{Kuhn2019}. (The full table is available
in the online journal. A portion is shown here for guidance regarding its form and content.)}
}
\end{table*}

\subsection{Data selection}
In order to ensure that the catalog contains only photometry of accreting stars, the source positions were matched with the catalog of accreting stars from \cite{Manara2012}, which are complete down to the hydrogen burning limit. This step reduced our sample size to $61$.
For $60$ sources, we found parallaxes in the Gaia Early Data Release 3 (EDR3) catalog \citep{Gaia2016b, Gaia2020b}. The mean parallax of the sources is $\left\langle \bar{\omega}\right\rangle = 2.48 \pm 0.06~\mathrm{mas}$, which corresponds to a mean distance of $\sim 403~\mathrm{pc}$, in agreement with the distance of $403^{+7}_{-6}~\mathrm{pc}$ determined by \cite{Kuhn2019}. We excluded three sources as likely background stars according to the criteria that their distance is larger than the one reported by \cite{Kuhn2019} within $3\sigma$ and kept the one source without a listed parallax value (\object{V1118 Ori}) and assigned to it the distance of $403~\mathrm{pc}$, with the typical uncertainty of $\approx \pm 9~\mathrm{pc}$ deduced from our sample.

We complemented the catalog with the effective temperatures determined by \cite{DaRio2012}. Where available, we took more recent spectral types obtained by \cite{Hillenbrand2013} and converted them to $T_\mathrm{eff}$ according to the temperature scales of \cite{Kenyon1995} and \cite{Luhman2003} for K and M type stars, respectively. During this process, we excluded one additional source due to its uncertain spectral classification. The final catalog contains data for $57$ sources. It is arranged in two tables: the first (Table \ref{tab:tabphot1}) lists the sources with the collected parameters from the literature. The second (Table \ref{tab:tabphot2}) comprises the photometric data from our survey and the epochs as well as the object names according to Table \ref{tab:tabphot1}.

\begin{table*}
\caption{\label{tab:tabphot2}WWFI $u^\prime g^\prime r^\prime$ photometry.}
\centering
\begin{tabular}{ccccc}
\hline\hline\\[-2.0ex]
Object\tablefootmark{(1)} & Julian date\tablefootmark{(2)} & \multicolumn{1}{c}{$u^\prime$}  & \multicolumn{1}{c}{$g^\prime$} & \multicolumn{1}{c}{$r^\prime$} \\
 &  $[$days$]$   & \multicolumn{1}{c}{$[\mathrm{mag}]$}    & \multicolumn{1}{c}{$[\mathrm{mag}]$}     & \multicolumn{1}{c}{$[\mathrm{mag}]$} \\[0.4ex]
\hline\\[-1.6ex]
\object{V397 Ori} & $ 2457730.6$ & $17.83 \pm 0.06$ & $15.85 \pm 0.08$ & $14.55 \pm 0.08$ \\
\object{V397 Ori} & $ 2457745.6$ & $18.12 \pm 0.06$ & $15.94 \pm 0.07$ & $14.61 \pm 0.07$ \\
\multicolumn{1}{c}{\vdots} & \multicolumn{1}{c}{\vdots} & \multicolumn{1}{c}{\vdots} & \multicolumn{1}{c}{\vdots} & \multicolumn{1}{c}{\vdots} \\[1ex]
\object{V1979 Ori} & $ 2457745.6$ & $19.27 \pm 0.07$ & $16.99 \pm 0.07$ & $15.57 \pm 0.07$ \\
\object{V1979 Ori} & $ 2457776.4$ & $19.56 \pm 0.05$ & $17.06 \pm 0.06$ & $15.62 \pm 0.06$ \\
\multicolumn{1}{c}{\ldots} & \multicolumn{1}{c}{\ldots} & \multicolumn{1}{c}{\ldots} & \multicolumn{1}{c}{\ldots} & \multicolumn{1}{c}{\ldots} \\[1ex] \hline
\end{tabular}
\tablefoot{
\tablefoottext{1}{Name of the YSO as listed in SIMBAD.}
\tablefoottext{2}{Epoch of the observation. (The full table is available
in the online journal. A portion is shown here for guidance regarding its form and content.)
}}
\end{table*}


\section{Method} \label{sec:method}

The accretion rates can be estimated from the observed $u^\prime$, $g^\prime$, and $r^\prime$ photometry and the effective temperatures known from the literature with an approach similar to the one used by \cite{DaRio2010} and \cite{Manara2012}, where the latter called it the ``2CD'' method.

The method assumes that the position of the sources in the $g^\prime - r^\prime$ versus $u^\prime - g^\prime$ color-color diagram depends solely on three parameters: the effective temperature, the accretion luminosity, and the extinction. Without accretion, the colors would fall on a line determined by the photosphere of the stars at a position set by the effective temperature and scatter along the reddening vector. Accretion shifts the position towards bluer values to the colors of a purely accreting source.

The accretion luminosity and the extinction can be obtained simultaneously from the observed magnitudes and the $T_\mathrm{eff}$ values when the photospheric colors and the colors of the accretion source are known. In the following, we describe our implementation of the approach in more detail and explain how we obtained the required photospheric colors.

\subsection{Flux density modeling} \label{sec:flux}

The flux density of the sources in question is assumed to be a composition of the pure photospheric flux density and the accretion flux density. Without extinction, the flux density of a source located at a distance $d$ can be expressed through
\begin{align}
F_\lambda & = I_{\mathrm{phot},\, \lambda} \cdot \frac{\pi R_*^2}{d^2} + I_{\mathrm{acc},\, \lambda} \cdot \frac{A_\mathrm{acc}}{d^2} \nonumber \\
          & = \left(\frac{R_*}{R_\odot}\right)^2 \cdot \left(F_{\mathrm{phot},\, \lambda} + \eta \cdot F_{\mathrm{acc},\, \lambda}\right), \label{eq:eqflux}
\end{align}
where $R_*$ is the radius of the source with the intensity $I_{\mathrm{phot},\, \lambda}$ and $A_\mathrm{acc}$ is the area of the accreting regions with the intensity $I_{\mathrm{acc},\, \lambda}$. The factor $(R_* / R_\odot)^2$ scales the fluxes according to the radius, where it is assumed that $F_{\mathrm{phot},\, \lambda}$ and $F_{\mathrm{acc},\, \lambda}$ refer to an area similar to $\pi {R_\odot}^2$, and $\eta \coloneqq A_\mathrm{acc} / (\pi R_*^2)$ describes the fraction of the projected area of the photosphere covered by the accreting regions.

\subsection{Color modeling} \label{sec:mag}
Using synthetic photometry (see \cite{Casagrande2014} for a detailed description of the topic), the flux density can be related to the expected magnitudes $m_\zeta$ for the $\zeta$ filter (where $\zeta$ stands for $u^\prime$, $g^\prime$, and $r^\prime$ or any other filter) via
\begin{equation}
m_\zeta = -2.5 \log \bar{F}_{\zeta} - Z_\zeta, \label{eq:eqmag}
\end{equation}
with the zero-point $Z_\zeta$ and
\begin{equation}
\bar{F}_\zeta = \frac{\int \lambda \, F_\lambda \, T_{\zeta, \lambda} \, \mathrm{d} \lambda}{\int \lambda \, T_{\zeta, \lambda} \, \mathrm{d} \lambda}, \label{eq:eqmflux}
\end{equation}
where $T_{\zeta, \lambda}$ is the passthrough function of the respective filter. The magnitude according to Eq. \ref{eq:eqmag} is given by
\begin{equation}
m_{\zeta} (\eta) = -2.5 \log \left(\bar{F}_{\mathrm{phot},\,\zeta} + \eta \cdot \bar{F}_{\mathrm{acc},\,\zeta}\right) - 5 \log\left(\frac{R_*}{R_\odot}\right) - Z_\zeta. \label{eq:eqmag2}
\end{equation}
The respective colors are independent of $R_*$:
\begin{equation}
m_{\zeta} (\eta) - m_{\mu} (\eta) = -2.5 \log \left(\frac{\bar{F}_{\mathrm{phot},\,\zeta} + \eta \cdot \bar{F}_{\mathrm{acc},\,\zeta}}{\bar{F}_{\mathrm{phot},\,\mu} + \eta \cdot \bar{F}_{\mathrm{acc},\,\mu}}\right) - Z_\zeta + Z_\mu. \label{eq:eqcolor}
\end{equation}
In the following subsections, we explain how we obtained the photospheric fluxes described by $\bar{F}_{\mathrm{phot},\,\zeta}$ and the accretion flux $\bar{F}_{\mathrm{acc},\,\zeta}$.

\subsubsection{Photospheric flux} \label{sec:methphototrack}

The photospheric flux $\bar{F}_{\mathrm{phot},\,\zeta}$ required for Eq. \ref{eq:eqcolor} was determined from empirically obtained spectra of young, diskless class III objects serving as photospheric templates. They were observed with X-Shooter by \cite{Manara2013} and \cite{Manara2017}. We excluded the spectra of the targets named \object{2MASS J11195652-7504529}, \object{[LES2004] ChaI 601}, \object{[LES2004] ChaI 717}, and \object{V1251 Cen} due to a low signal-to-noise ratio.

For each of the remaining template spectra, we calculated $F_{\mathrm{phot},\, \lambda}$ according to Eq. \ref{eq:eqflux} from the observed flux density by scaling the distance of the template source, taken from the latest Gaia EDR3 release \citep{Gaia2016b, Gaia2020b}, to the distance of the ONC. Furthermore, the radius of the template source was scaled to the solar radius. We obtained the radius of the template source from the luminosities determined by \cite{Manara2013, Manara2017}.

With $F_{\mathrm{phot},\, \lambda}$, we calculated $\bar{F}_{\mathrm{phot},\,\zeta}$ with Eq. \ref{eq:eqmflux} using the passthrough functions $T_{\zeta, \lambda}$ of the $u^\prime$, $g^\prime$, and $r^\prime$ filters used by the WWFI setup \citep{Kosyra2014}. In order to increase the signal-to-noise ratio, the template spectra were median smoothed with a typical box-width of $\sim 3~\mbox{\AA}$. For convenience, we converted the fluxes to magnitudes using Eq. \ref{eq:eqmag} and denoted them $m_{\mathrm{phot},\, \zeta}$.

Since the relation between these magnitudes and $T_\mathrm{eff}$ is complicated and to avoid a possible bias by assuming a specific shape, we utilized a nonparametric regression approach in order to approximate the relation between the magnitudes and the effective temperature. Similar to \cite{Manara2017}, we used a local second degree polynomial regression with a Gaussian kernel implemented in the Python toolkit \verb|pyqt_fit.npr_methods.LocalPolynomialKernel| for this task. The $2\sigma$ confidence level was obtained using bootstrapping. 

In the final step, the fitting results were sampled at $20$ $T_\mathrm{eff}$ values, from which the fitting results of the nonparametric regression can be recreated by interpolating with cubic splines. This allows $\bar{F}_{\mathrm{phot},\,\zeta}(T_\mathrm{eff})$ to be obtained for every given $T_\mathrm{eff}$. A plot of the relation between $m_{\mathrm{phot},\, \zeta}$ and $T_\mathrm{eff}$ as well as the fitting results are shown in Appendix \ref{app:iso}.

\subsubsection{The accretion colors} \label{sec:methaccretion}

In order to compare Eq. \ref{eq:eqcolor} with the observed data, the flux associated with the accretion $\bar{F}_{\mathrm{acc},\,\zeta}$ is required. We chose the models of \cite{Manara2013}, which consist of the intensity of a slab of pure hydrogen (bound-free and free-free emission from H and H$^-$) assuming local thermodynamic equilibrium. The models depend on three parameters: the temperature ($T_\mathrm{slab}$), the optical depth at a reference wavelength of $\lambda = 300~\mathrm{nm}$ ($\tau_{300}$), and the electron density ($n_{\rm e}$).

We calculated the flux density $F_{\mathrm{acc},\,\lambda}$ according to Eq. \ref{eq:eqflux} from the intensity $I_{\mathrm{acc},\,\lambda}$ of the slab and obtained the magnitudes in the WWFI filters for several model parameters using Eq. \ref{eq:eqmag} and Eq. \ref{eq:eqmflux}. We denoted them $m_{\mathrm{acc},\,\zeta}$. In addition, we determined the luminosity $L_{\mathrm{acc},\,0}$ of the model slab by integrating the flux density and setting a distance of $d = 403~\mathrm{pc}$.

\begin{figure}
\resizebox{\hsize}{!}
        {
                \includegraphics[width=\hsize]{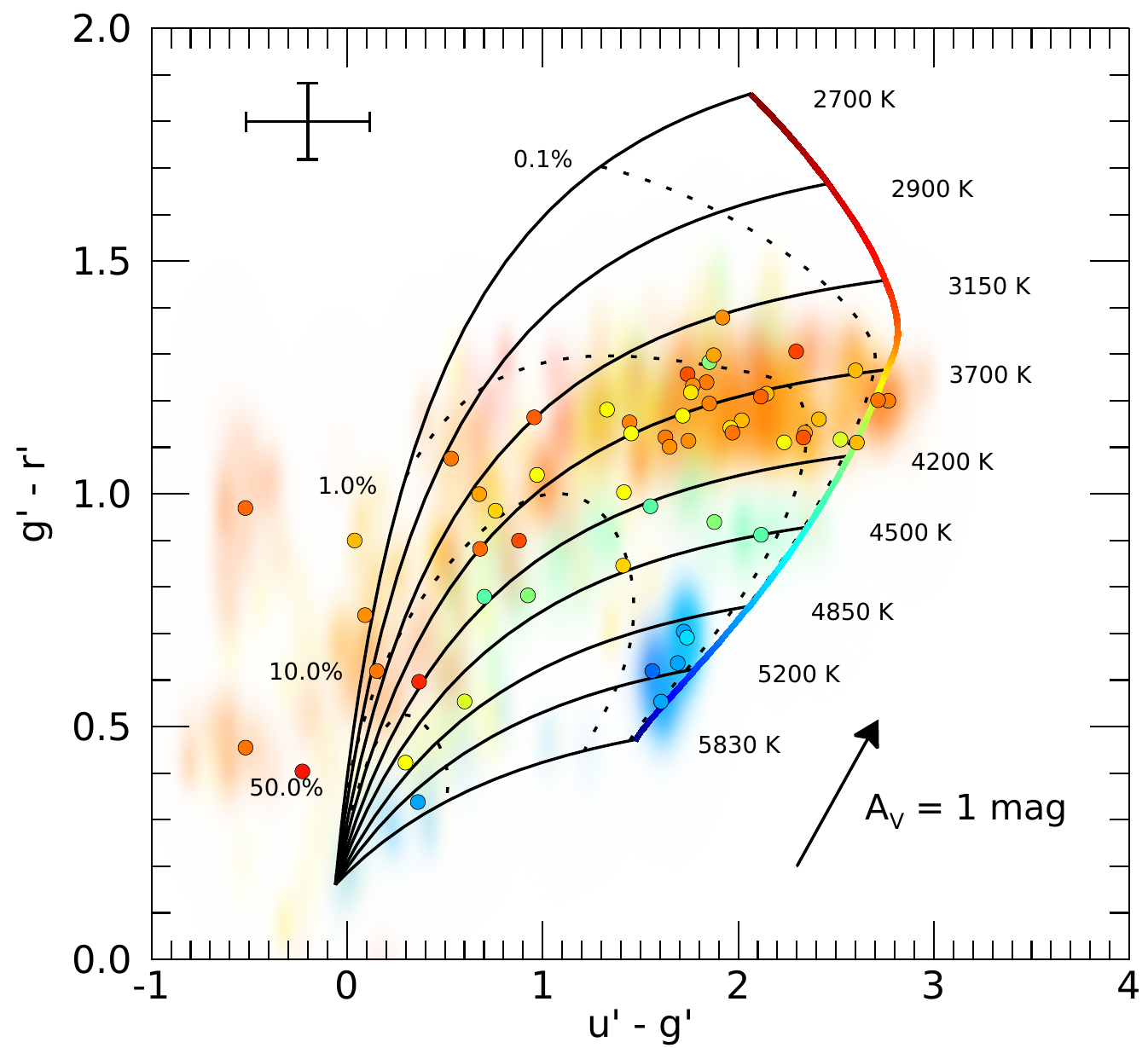}
        }
        \caption{Color-color diagram of the WWFI sample, dereddened with the extinction values determined by \cite{Manara2012} and plotted as 2D gaussians according to their photometric uncertainties. The values are colored according to their $T_\mathrm{eff}$ value. Overlaid are the mean colors for each star as solid dots. The cross in the upper left corner indicates their typical range of variability. The thick line shows the photospheric colors for non-accreting stars. The thin lines represent tracks along which the sources are displaced from the photospheric track for increasing accretion. They converge in a point representing the pure accretion spectrum, described by a hydrogen slab model. The dotted lines indicate constant accretion to total luminosities ratios, $L_\mathrm{acc} / L_\mathrm{tot}$.} \label{fig:fig2cd2}
\end{figure}

In order to select an appropriate accretion model for our data, we dereddened the colors of our sample with the extinction values determined by \cite{Manara2012}. The resulting color-color diagram is shown in Fig. \ref{fig:fig2cd2} and illustrates how the colors are distributed when neglecting extinction. Furthermore, Fig. \ref{fig:fig2cd2} shows the predicted displacement of the sources from the line describing the photosphere according to Eq. \ref{eq:eqcolor}. For a given T$_\mathrm{eff}$ value, a track in the color-color diagram is determined by varying $\eta$. The track starts for $\eta = 0$ at the line describing the pure photospheric emission and converges at the colors of the pure accretion spectrum for increasing $\eta$. The plot also shows lines of constant ratio between the accretion luminosity and the total luminosity, $L_\mathrm{acc} / L_\mathrm{tot}$. The calculation of these lines is detailed in appendix \ref{app:lines}.

The model with the parameters $T_\mathrm{slab} = 11000~\mathrm{K}$, $\tau_{300} = 5.00$, and $n_{\rm e} = 10^{15}~\mathrm{cm}^{-3}$ lead to tracks describing the extinction-corrected colors best. The corresponding magnitudes are $m_{\mathrm{acc},\,u^\prime} = 10.70~\mathrm{mag}$, $m_{\mathrm{acc},\,g^\prime} = 10.76~\mathrm{mag}$ and $m_{\mathrm{acc},\,r^\prime} = 10.60~\mathrm{mag}$, while the accretion luminosity amounts to $\log(L_{\mathrm{acc},\,0} / L_\odot) = 1.04$. We use these parameters for our subsequent analysis.

\subsection{Accretion luminosities and extinctions from observations} \label{sec:lum}

We applied Eq. \ref{eq:eqcolor} to each $u^\prime g^\prime r^\prime$ observation and derived $\eta$ and $A_V$ using $T_\mathrm{eff}$, the photospheric flux, and the accretion model. The observed position in the color-color diagram is shifted along the reddening vector, until it intersects the unique track calculated from $T_\mathrm{eff}$. This procedure provides $A_V$ as well as $\eta$. The yet missing radius $R_*$ can be calculated from the observed magnitudes, $A_V$, and $\eta$ with Eq. \ref{eq:eqmag2}. From the definition of $\eta$, the accretion luminosity follows from
\begin{equation}
L_\mathrm{acc} = \eta \cdot \left(\frac{R_*}{R_\odot}\right)^2 \cdot L_{\mathrm{acc},\,0}. \label{eq:eqlacc}
\end{equation}
In the next subsection, we explain in more detail how we estimated the uncertainties of $A_V$, $R_*$, and $L_\mathrm{acc}$.

\subsection{Uncertainties} \label{sec:secerr}

In order to estimate the uncertainties, we used a Monte Carlo approach similar to \cite{Manara2012}: the colors were randomly displaced according to their photometric uncertainty assuming Gaussian. The same was done with $T_\mathrm{eff}$, assuming an uncertainty corresponding to one stellar subclass. A limitation of our approach is the possibility that the spectral types of the stars in our sample changed by more than one stellar subclass in the time span between their determination and our photometric monitoring.

The $1\sigma$ uncertainty of the photospheric magnitudes is assumed to be the $1\sigma$ confidence interval. The accretion luminosity and the extinction were then determined with the method described above. This step is repeated $10^4$ times and the $1\sigma$ uncertainty is defined as the standard deviation of the obtained set of values. With this method, we estimated uncertainties for $A_V$, $R_*$, and $L_\mathrm{acc}$.

If there was more than one solution, that is the line intersects the track at more than one point, we chose the solution with the lower accretion luminosity. For some iterations, no solution could be found. This is the case when the colors are not inside the range of the combination of the model accretion spectrum and the extinction. Similar to \cite{Manara2012}, we assigned the results a confidence level according to the amount
of successful intersections: if there was a successful intersection for all $10^4$ iterations, we assigned the result a confidence level of $3\sigma$ and a lower confidence level of $2\sigma$ and $1\sigma$ if there were only successful intersections for more than $95~\%$ and $68~\%$ of the iterations, respectively.

For the further analysis, we only used sources with a confidence level $\geq 1\sigma$ and $A_V \geq 0~\mathrm{mag}$. Furthermore, we limited ourselves to stars for which at least five epochs are available.

\subsection{Accretion rates and stellar parameters} \label{sec:parameters}
The $A_V$ values obtained with our method range from $0.00~\mathrm{mag}$ to $3.04~\mathrm{mag}$, with a mean value of $0.70~\mathrm{mag}$. For each star, we calculated the mean of the extinction values and compared the results with the extinctions obtained by \cite{Manara2012}. We found a good match based on the distribution of the differences $\Delta A_\mathrm{V} = \left\langle A_{\mathrm{V,\,this\,work}}\right\rangle - A_{\mathrm{V,\,Manara}}$, which displays a median value of $-0.09~\mathrm{mag}$ and a standard deviation of $0.57~\mathrm{mag}$.

\begin{figure}
\resizebox{\hsize}{!}
        {
                \includegraphics[width=\hsize]{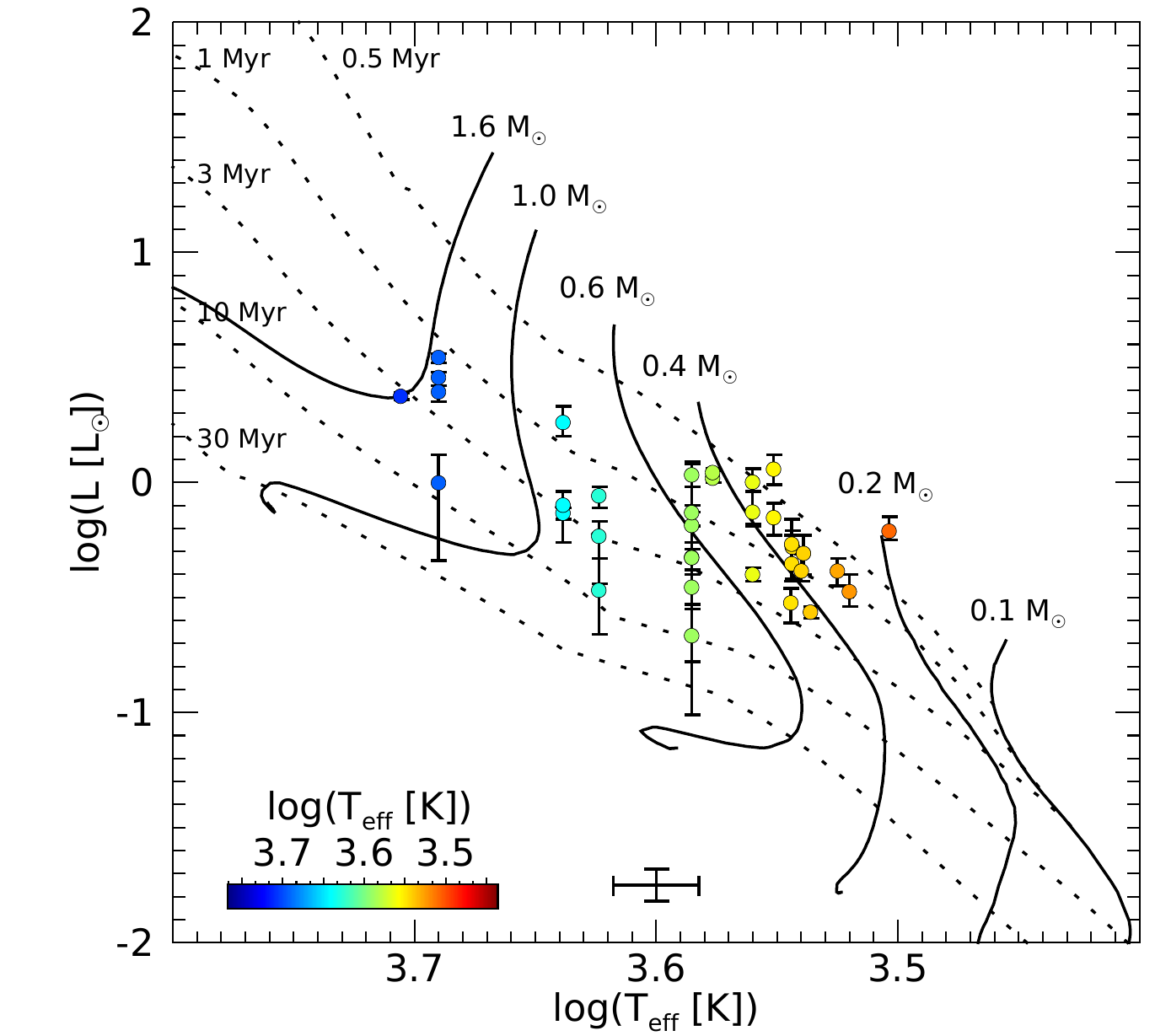}
        }
        \caption{Hertzsprung-Russel diagram of the observed sources. Overlaid are the PARSEC 1.2S isochrones and mass tracks from \cite{Bressan2012}.
				The dots represent the mean bolometric luminosity for each star as a function of $T_\mathrm{eff}$. The vertical bars indicate
				the minimum and maximum range of $L_*$ for each star. The cross at the bottom shows the typical uncertainty.
				In our analysis, we only regarded sources with $0.2~M_\odot \leq M \leq 2.0~M_\odot$ and $5.5 \leq \log(\tau [\mathrm{yr}]) \leq 7.3$ in order to exclude outliers in the Hertzsprung-Russel diagram.				
				} \label{fig:fighr}
\end{figure}

With the given $T_\mathrm{eff}$ values and the radii obtained with our method, the bolometric luminosities $L_*$ can be calculated via $L_* / L_\odot = (R_* / R_\odot)^2 \cdot (T_\mathrm{eff} / T_\odot)^4$. From the bolometric luminosities and the known spectral types, we estimated the stellar masses $M$ and ages $\tau$ by interpolating PARSEC 1.2S isochrones and mass tracks from \cite{Bressan2012} on the Hertzsprung-Russel diagram (HRD), shown in Fig. \ref{fig:fighr}. Under the assumption that the gravitational infall releases its energy at a magnetospheric radius of $\sim 5$ stellar radii \citep{Shu1994}, the accretion rate can be estimated from $L_\mathrm{acc}$ \citep[e.g.,][]{Hartmann2016}:
\begin{equation}
	\dot{M}_\mathrm{acc} \approx \frac{L_\mathrm{acc} \, R_*}{0.8\,G\,M}. \label{eq:eqmacc}
\end{equation}
Similar to Sec. \ref{sec:secerr}, the uncertainties of the masses, ages, and accretion rates were estimated with a Monte Carlo approach, propagating the uncertainties from the photometry, the effective temperatures, and the photospheric magnitudes. The results are summarized in Table \ref{tab:tabparams}.

\begin{figure}
\resizebox{\hsize}{!}
        {
                \includegraphics[width=\hsize]{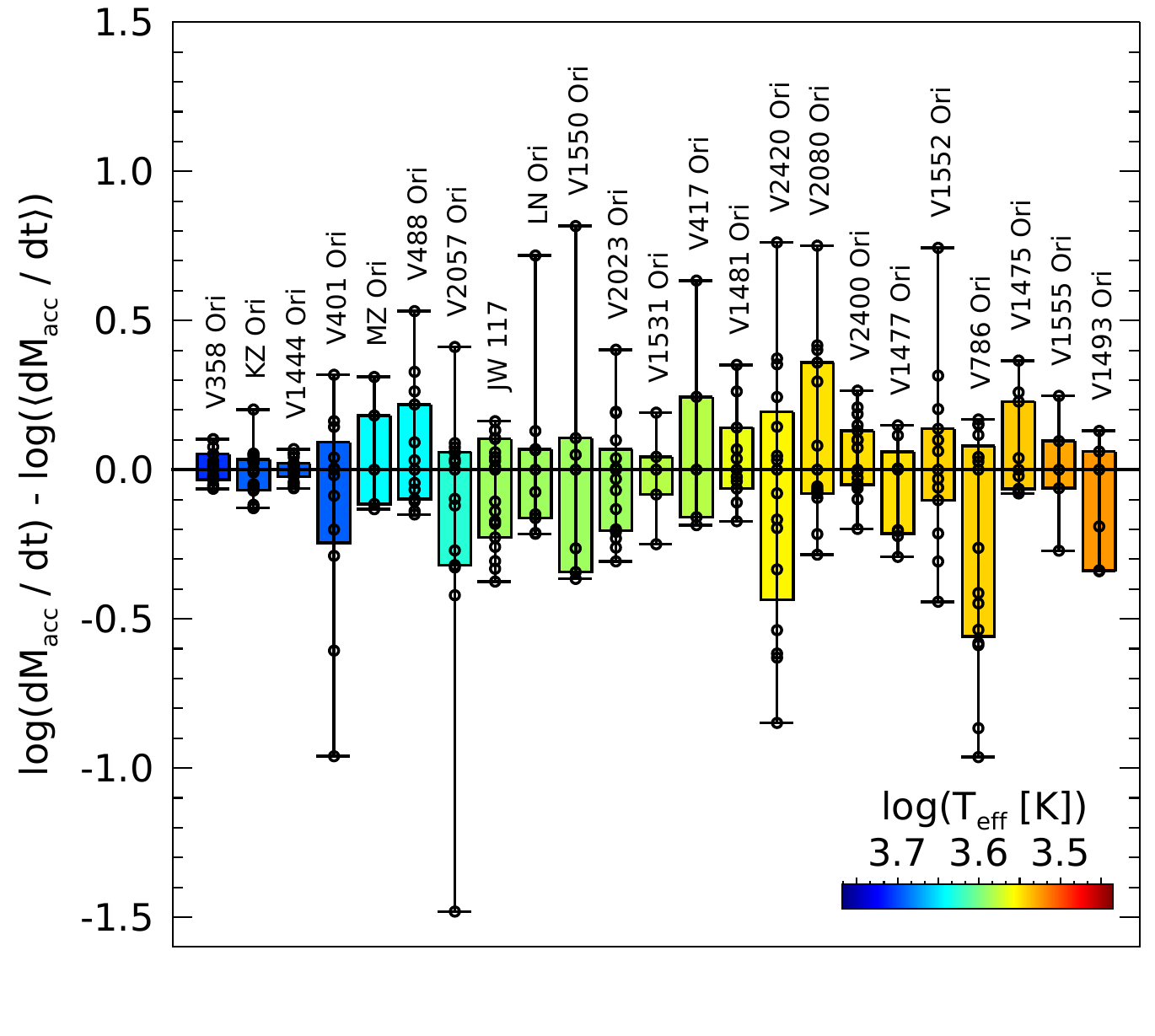}
        }
        \caption{Boxplot showing the accretion rate variability of the targets. The vertical solid lines indicate the range of the accretion rates and the height of the boxes the interquartile range. The circles show the accretion rates for each target and epoch. The values are normalized to the median values and color-coded according to their effective temperature.
				} \label{fig:figboxplot}
\end{figure} 

\subsection{The final sample}

For the further analysis, we calculated the ratio $L_\mathrm{acc} / L_*$ and compared it with relation (1) derived by \cite{Manara2017} in order to exclude values that are likely below the ``noise level'' due to photospheric emission. In addition, we limited the sample according to $0.2~M_\odot \leq M \leq 2.0~M_\odot$ and $5.5 \leq \log(\tau [\mathrm{yr}]) \leq 7.3$ in order to exclude outliers in the HRD.

Since we are primarily interested in the influence of the accretion rate variability on the relation between X-ray emission and accretion, we limited our sample further to stars with a match with the X-ray data from the COUP catalog \citep{Getman2005}. This left us with a total of $23$ stars with $\sim 12$ mass accretion rate values per average, good $u^\prime g^\prime r^\prime$ photometry for $\geq 5$ nights, Gaia distances consistent with the ONC, evidence of accretion, and a confidence level $\geq 1\sigma$. For each star, we also obtained the mean values of the stellar parameters and propagated their uncertainties. We report the resulting values in Table \ref{tab:tabvar}.

\begin{table*}
\caption{\label{tab:tabparams}Stellar parameters and accretion values of the WWFI sample.}
\centering
\begin{tabular}{ccccccc}
\hline\hline\\[-2.0ex]
Object\tablefootmark{(1)} & Julian date\tablefootmark{(2)} & \multicolumn{1}{c}{$A_V$}  & \multicolumn{1}{c}{$\log(L_\mathrm{acc})$} & \multicolumn{1}{c}{$\log(L_*)$} & \multicolumn{1}{c}{$\log(\dot{M}_\mathrm{acc})$} & \multicolumn{1}{c}{Confidence level\tablefootmark{(3)}}
\\
 &  $[$days$]$   & \multicolumn{1}{c}{$[\mathrm{mag}]$} & \multicolumn{1}{c}{$(L_\odot)$} & \multicolumn{1}{c}{$(L_\odot)$} &
\multicolumn{1}{c}{$(M_\odot / \mathrm{yr})$} & \multicolumn{1}{c}{$[\sigma]$} \\[0.4ex]
\hline\\[-1.6ex]
\object{V397 Ori} & $2457730.6$ & $1.03 \pm 0.26$ & $-1.31 \pm 0.15$ & $-0.21 \pm 0.02$ & $-8.43 \pm 0.16$ & $3$ \\ 
\object{V397 Ori} & $2457745.6$ & $1.05 \pm 0.26$ & $-1.49 \pm 0.16$ & $-0.22 \pm 0.02$ & $-8.62 \pm 0.17$ & $3$ \\ 
\multicolumn{1}{c}{\vdots} & \multicolumn{1}{c}{\vdots} & \multicolumn{1}{c}{\vdots} & \multicolumn{1}{c}{\vdots} & \multicolumn{1}{c}{\vdots} & \multicolumn{1}{c}{\vdots} & \multicolumn{1}{c}{\vdots} \\[1ex]
\object{V976 Ori} & $2457730.6$ & $0.12 \pm 0.38$ & $-1.95 \pm 0.23$ & $-0.14 \pm 0.04$ & $-9.12 \pm 0.24$ & $2$ \\ 
\object{V976 Ori} & $2457745.6$ & $0.13 \pm 0.39$ & $-2.08 \pm 0.25$ & $-0.16 \pm 0.04$ & $-9.26 \pm 0.26$ & $2$ \\ 
\multicolumn{1}{c}{\ldots} & \multicolumn{1}{c}{\ldots} & \multicolumn{1}{c}{\ldots} & \multicolumn{1}{c}{\ldots} & \multicolumn{1}{c}{\ldots} & \multicolumn{1}{c}{\ldots} & \multicolumn{1}{c}{\ldots} \\[1ex]
\hline
\end{tabular}
\tablefoot{
\tablefoottext{1}{Name of the YSO as listed in SIMBAD.}
\tablefoottext{2}{Epoch of the observation.}
\tablefoottext{3}{Confidence level as defined in Sec. \ref{sec:secerr}. (The full table is available
in the online journal. A portion is shown here for guidance regarding its form and content.)}
}
\end{table*}

\begin{table*}
\caption{\label{tab:tabvar}Stellar parameters and accretion variability.}
\centering
\begin{tabular}{lrcccccr}
\hline\hline\\[-2.0ex]
Object\tablefootmark{(1)} & \multicolumn{1}{c}{Epochs\tablefootmark{(2)}} & $M$ & \multicolumn{1}{c}{$\log(\tau)$} & \multicolumn{1}{c}{$\log(\langle\dot{M}_\mathrm{acc}\rangle)$\tablefootmark{(3)}} & 
\multicolumn{1}{c}{$\Delta\log(\dot{M}_\mathrm{acc})$\tablefootmark{(4)}} & \multicolumn{1}{c}{$\Delta_\mathrm{IQR}\log(\dot{M}_\mathrm{acc})$\tablefootmark{(5)}} & \multicolumn{1}{c}{COUP\tablefootmark{(6)}}
\\
 & & \multicolumn{1}{c}{$[M_\odot]$} & \multicolumn{1}{c}{$(\mathrm{yr})$} & \multicolumn{1}{c}{$(M_\odot / \mathrm{yr})$} & \multicolumn{1}{c}{$[\mathrm{dex}]$} & \multicolumn{1}{c}{$[\mathrm{dex}]$} & \\[0.4ex]
\hline\\[-1.6ex]
\object{V358 Ori}  & 15 & $1.57 \pm 0.02$ & $6.54 \pm 0.02$ & $-8.13 \pm 0.05$ & $0.17 \pm 0.40$ & $0.09 \pm 0.10$ & 1269 \\
\object{KZ Ori}    & 20 & $1.46 \pm 0.04$ & $6.33 \pm 0.02$ & $-8.38 \pm 0.05$ & $0.33 \pm 0.47$ & $0.10 \pm 0.10$ & 188 \\
\object{V1444 Ori} & 19 & $1.51 \pm 0.05$ & $6.13 \pm 0.03$ & $-7.96 \pm 0.05$ & $0.13 \pm 0.43$ & $0.04 \pm 0.09$ & 23 \\
\object{V401 Ori}  & 12 & $1.24 \pm 0.02$ & $6.86 \pm 0.06$ & $-7.44 \pm 0.06$ & $1.28 \pm 0.30$ & $0.34 \pm 0.14$ & 62 \\
\object{MZ Ori}    &  5 & $0.88 \pm 0.05$ & $6.38 \pm 0.06$ & $-7.61 \pm 0.14$ & $0.44 \pm 0.41$ & $0.30 \pm 0.21$ & 1134 \\
\object{V488 Ori}  & 13 & $0.79 \pm 0.03$ & $5.89 \pm 0.02$ & $-7.60 \pm 0.08$ & $0.68 \pm 0.47$ & $0.32 \pm 0.15$ & 567 \\
\object{V2057 Ori} & 15 & $0.80 \pm 0.02$ & $6.97 \pm 0.03$ & $-8.44 \pm 0.06$ & $1.89 \pm 0.39$ & $0.38 \pm 0.12$ & 54 \\
\object{JW 117}    & 19 & $0.49 \pm 0.02$ & $5.77 \pm 0.01$ & $-7.64 \pm 0.05$ & $0.54 \pm 0.35$ & $0.33 \pm 0.09$ & 58 \\
\object{LN Ori}    & 10 & $0.65 \pm 0.02$ & $6.40 \pm 0.01$ & $-8.25 \pm 0.06$ & $0.93 \pm 0.32$ & $0.23 \pm 0.11$ & 301 \\
\object{V1550 Ori} &  7 & $0.56 \pm 0.03$ & $6.06 \pm 0.02$ & $-7.95 \pm 0.15$ & $1.18 \pm 0.36$ & $0.45 \pm 0.16$ & 1421 \\
\object{V2023 Ori} & 16 & $0.58 \pm 0.02$ & $6.11 \pm 0.01$ & $-8.43 \pm 0.05$ & $0.71 \pm 0.28$ & $0.30 \pm 0.08$ & 28 \\
\object{V1531 Ori} &  5 & $0.44 \pm 0.03$ & $5.72 \pm 0.01$ & $-8.36 \pm 0.07$ & $0.44 \pm 0.21$ & $0.13 \pm 0.12$ & 1248 \\
\object{V417 Ori}  &  5 & $0.43 \pm 0.03$ & $5.68 \pm 0.02$ & $-7.81 \pm 0.11$ & $0.82 \pm 0.25$ & $0.40 \pm 0.16$ & 1333 \\
\object{V1481 Ori} & 11 & $0.36 \pm 0.02$ & $5.59 \pm 0.02$ & $-8.57 \pm 0.04$ & $0.52 \pm 0.22$ & $0.20 \pm 0.09$ & 202 \\
\object{V2420 Ori} & 16 & $0.34 \pm 0.01$ & $5.78 \pm 0.02$ & $-8.26 \pm 0.06$ & $1.61 \pm 0.25$ & $0.58 \pm 0.12$ & 1282 \\
\object{V2080 Ori} & 14 & $0.47 \pm 0.01$ & $6.28 \pm 0.02$ & $-9.21 \pm 0.05$ & $1.04 \pm 0.23$ & $0.44 \pm 0.10$ & 112 \\
\object{V2400 Ori} & 17 & $0.34 \pm 0.01$ & $5.87 \pm 0.02$ & $-8.65 \pm 0.03$ & $0.46 \pm 0.24$ & $0.18 \pm 0.07$ & 1236 \\
\object{V1477 Ori} &  8 & $0.35 \pm 0.02$ & $5.91 \pm 0.02$ & $-8.63 \pm 0.05$ & $0.44 \pm 0.23$ & $0.32 \pm 0.11$ & 139 \\
\object{V1552 Ori} & 13 & $0.36 \pm 0.02$ & $5.99 \pm 0.02$ & $-8.93 \pm 0.07$ & $1.19 \pm 0.29$ & $0.24 \pm 0.10$ & 1432 \\
\object{V786 Ori}  & 16 & $0.33 \pm 0.01$ & $5.91 \pm 0.02$ & $-8.45 \pm 0.04$ & $1.13 \pm 0.23$ & $0.65 \pm 0.10$ & 179 \\
\object{V1475 Ori} &  9 & $0.43 \pm 0.02$ & $6.29 \pm 0.03$ & $-9.27 \pm 0.05$ & $0.45 \pm 0.22$ & $0.29 \pm 0.10$ & 132 \\
\object{V1555 Ori} &  5 & $0.28 \pm 0.02$ & $5.84 \pm 0.03$ & $-8.42 \pm 0.10$ & $0.52 \pm 0.32$ & $0.16 \pm 0.17$ & 1454 \\
\object{V1493 Ori} &  6 & $0.28 \pm 0.02$ & $6.01 \pm 0.02$ & $-8.94 \pm 0.09$ & $0.47 \pm 0.34$ & $0.40 \pm 0.18$ & 546
  \\\hline
\end{tabular}
\tablefoot{
\tablefoottext{1}{Name of the YSO as listed in SIMBAD. The rows are sorted by descending $T_\mathrm{eff}$ values.}
\tablefoottext{2}{The mean baseline covered by the epochs amounts to $\sim 755$ days.}
\tablefoottext{3}{Mean value of the accretion rates.}
\tablefoottext{4}{Variability range defined as $\Delta\log(\dot{M}_\mathrm{acc}) = \log{(\dot{M}_{\mathrm{acc,\,max}})} - \log{(\dot{M}_{\mathrm{acc,\,min}})}$.}
\tablefoottext{5}{Interquartile range defined as the difference between the upper and the lower quartile of the logarithmic accretion rates.}
\tablefoottext{6}{COUP source number as listed in J/ApJS/160/319/coup \citep{Getman2005}.}
}
\end{table*}

\section{Results} \label{sec:secresults}

   \subsection{Accretion rate variability}
In order to probe the variability of the obtained accretion rates, we calculated the range of variability of the accretion rates $\Delta\log\left(\dot{M}_\mathrm{acc}\right)$, defined as the difference between the maximum and the minimum logarithmic accretion rate for each target. In addition, we calculated the interquartile range (IQR) $\Delta_\mathrm{IQR}\log\left(\dot{M}_\mathrm{acc}\right)$, which is defined as the difference between the upper and the lower quartile.

To estimate the uncertainties, we used again a Monte Carlo approach: the mass accretion rates were varied according to their uncertainty assuming Gaussian and $\Delta\log\left(\dot{M}_\mathrm{acc}\right)$ as well as $\Delta_\mathrm{IQR}\log\left(\dot{M}_\mathrm{acc}\right)$ was calculated for each source. This step was repeated $10^4$ times and the uncertainty estimated as the $1\sigma$ standard deviation drawn from the obtained distributions. The resulting values are listed in Table \ref{tab:tabvar}.

Figure \ref{fig:figboxplot} shows a box plot of the mass accretion rates for each source, subtracted by their median. The range of the variability varies between $0.13~\mathrm{dex}$ and $1.89~\mathrm{dex}$, with a median value of $\sim 0.54~\mathrm{dex}$. We estimated the uncertainty of the median variability with the same method as described above and obtained $\sim 0.10~\mathrm{dex}$. The accretion rate variability is lower than the scatter found in \citetalias{Flaischlen2021} for the $\dot{M}_\mathrm{acc} - M$ relation of $0.80~\mathrm{dex}$, defined as the standard deviation of the residua obtained by subtracting the regression line from the observed $\dot{M}_\mathrm{acc}$ values. The values agree within $1\sigma$ with the typical mass accretion rate variability of $\sim 0.5~\mathrm{dex}$ determined for week-timescales \citep[e.g.,][]{Venuti2014} and is lower than the value of $\sim 0.65~\mathrm{dex}$ in the time range of several months found by \cite{Nguyen2009} using H$\alpha$ equivalent widths. Our results support previous findings that the major contribution to the accretion variability is found on shorter timescales than the year-timescale. 
The IQR varies between $0.04~\mathrm{dex}$ and $0.65~\mathrm{dex}$, with a median value of $0.30~\mathrm{dex}$ and a typical uncertainty of $\sim 0.04~\mathrm{dex}$.  

A similar consideration regarding the extinction yields $\Delta_\mathrm{IQR}A_V = (0.22 \pm 0.04)~\mathrm{mag}$, a variability compatible within $1\sigma$ with the variability of the mass accretion rates. The IQR variability of the bolometric luminosity reads $\Delta_\mathrm{IQR} \log(L_* / L_\odot) = 0.06 \pm 0.01$, in accordance with the typical uncertainty of $\sim 0.07~\mathrm{dex}$ we obtained for the (logarithmic) bolometric luminosity with our Monte Carlo method.
We also calculated the IQR variability of the obtained masses and found $\Delta_\mathrm{IQR}M = (0.02 \pm 0.01)~M_\odot$.

\begin{figure}
\resizebox{\hsize}{!}
        {
                \includegraphics[width=\hsize]{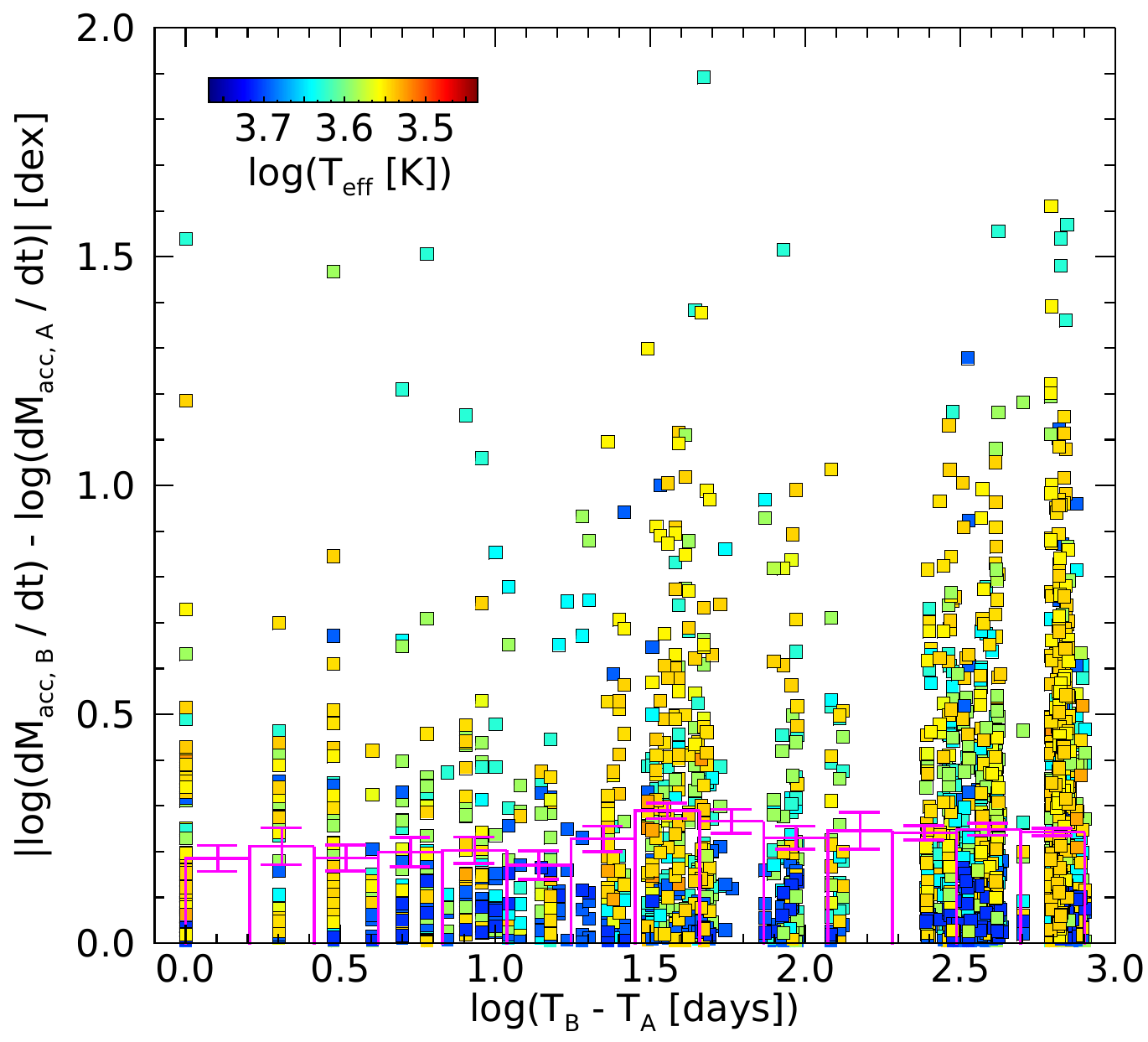}
        }
        \caption{Differences of the logarithmic mass accretion rates for two epochs $T_A$ and $T_B$ as a function of the logarithmic time differences $T_B - T_A$ for each measurement. The values are plotted as squares, where for each distinct object a different color was chosen according to its effective temperature. Overlaid are the mean values of the logarithmic mass accretion rate differences for different time bins as magenta columns, together with their estimated $1\sigma$ uncertainties.} \label{fig:figtimescales}
\end{figure}

In order to probe the accretion rate variability for all the timescales available in our sample, we followed the approach of \cite{Costigan2012, Costigan2014} and calculated for each object the difference between each observation timestamp and all other observation timestamps. Then, we plotted the respective mass accretion rate differences as a function of the time differences. The result is shown in Fig. \ref{fig:figtimescales}. Furthermore, we determined the mean of the mass accretion rate differences for bins of equal logarithmic duration and propagated the uncertainties. Although the data are irregularly and undersampled, one can deduce that the majority of the mass accretion rate variability is given by the days-to-weeks timescale.

\subsection{Relation between accretion rates and stellar mass}

We performed a linear regression in order to analyze the relation between the (logarithmic) mean masses and the mean (logarithmic) mass accretion rates. To this aim, we used the fully Bayesian \verb|LINMIX_ERR| approach based on the method developed by \cite{Kelly2007} that takes uncertainties in both variables into account as well as intrinsic scatter. The resulting relation reads
\begin{equation}
\log\left(\frac{\langle \dot{M}_\mathrm{acc} \rangle}{M_\odot \, \mathrm{yr}^{-1}}\right) = (-8.05 \pm 0.15) + (1.07 \pm 0.43) \cdot \log\left(\frac{M}{M_\odot}\right), \label{eq:eqmaccm}
\end{equation}
with a linear correlation coefficient of $r = 0.51 \pm 0.19$, indicating a positive correlation and in agreement with the slope of $1.07 \pm 0.22$ we found in our analysis of $332$ accreting sources in \citetalias{Flaischlen2021}.

\begin{figure}
\resizebox{\hsize}{!}
        {
                \includegraphics[width=\hsize]{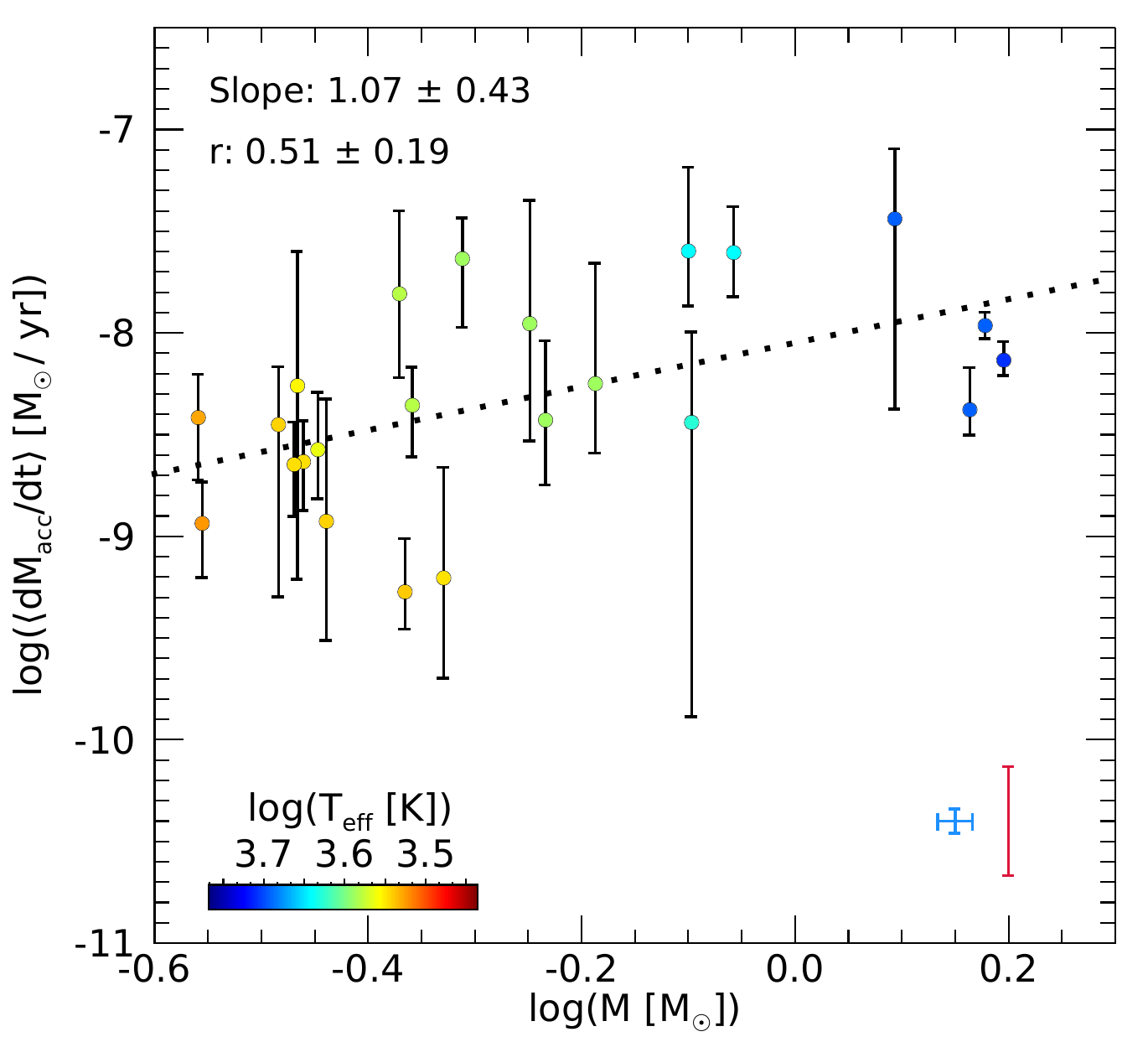}
        }
        \caption{Logarithmic mean mass accretion rates vs. logarithm of the masses shown as solid dots. The solid bars indicate the maximum and the minimum accretion rates for each source. The dotted line shows the result of the regression obtained with the Bayesian LINMIX\_ERR method. The blue cross shows the typical uncertainty of the values, the red bar the typical variability range.} \label{fig:figmaccm}
\end{figure}

In Fig. \ref{fig:figmaccm}, the regression result is displayed as a dotted line. We assume that the agreement of the regression results from our smaller sample with the results of the larger sample used in \citetalias{Flaischlen2021} indicates that it is not strongly affected by selection effects and therefore representative for the population of accreting stars in the ONC.

\subsection{Relation between X-ray activity and accretion}

The main goal of this work is to check whether the accretion variability introduces a bias in the detected anticorrelation between X-ray luminosities $L_X$ and mass accretion rates $\dot{M}_\mathrm{acc}$, which can be interpreted as a signature of X-ray driven photoevaporation. Under the assumption that the mean accretion rates over the timescale of $\sim 2$ years are more characteristic than a single ``snapshot'' of the accretion rates obtained during a single observation, we repeated the analysis using the mean values. We utilized a partial regression analysis for this task. The method is described in more detail in \citetalias{Flaischlen2021} and can be summarized as follows: since the accretion rates are correlated with the stellar mass (as was shown in numerous works, e.g., \cite{Alcala2017} and the previous section) and the X-ray luminosities as well \citep[e.g.,][]{Preibisch2005,Telleschi2007}, the values must be corrected for the common mass dependence in order to avoid spurious correlations. This is done by obtaining the linear regression result of the $\dot{M}_\mathrm{acc} - M$ and the $L_\mathrm{X} - M$ relations and calculating the residuals. If there was no intrinsic relation between $L_X$ and $\dot{M}_\mathrm{acc}$, the linear regression of the residuals would indicate no correlation.

To minimize the effect of X-ray variability, we tried to correct the influence of flaring activity from the X-ray luminosities using the ``characteristic count rates'' determined by \cite{Wolk2005} as the quiescent levels of X-ray emission outside the time windows with significant flaring activity in the light curves of the COUP sources. The corrected values are denoted ``characteristic'' X-ray luminosities, $L_\mathrm{X,\,char}$. Additionally, we corrected the values for the more recent Gaia distances listed in Table \ref{tab:tabphot1}. We assumed a typical uncertainty of $L_\mathrm{X,\,char}$ of $0.15~\mathrm{dex}$ derived from the spectral fits used to obtain the X-ray luminosities \citep{Preibisch2005}. Next, we performed the linear regression with \verb|LINMIX_ERR| and found the relation
\begin{equation}
\log\left(\frac{L_\mathrm{X,\,char}}{\mathrm{erg}\,\mathrm{s}^{-1}}\right) = (30.33 \pm 0.15) + (1.54 \pm 0.45) \cdot \log\left(\frac{M}{M_\odot}\right), \label{eq:eqlxm}
\end{equation}
with a linear correlation coefficient of $r = 0.65 \pm 0.16$, indicating a significant positive correlation. The slope is flatter than the one obtained in our previous analysis with the larger sample ($2.08 \pm 0.16$). 

We calculated the residuals using Equations \ref{eq:eqmaccm} and \ref{eq:eqlxm}, propagated the uncertainties and performed the linear regression with \verb|LINMIX_ERR|. We found the following relation:
\begin{equation}
\log\left(\frac{\langle\dot{M}_\mathrm{acc}\rangle}{\dot{M}_\mathrm{acc}(M)}\right) = 0.00 \pm 0.11 + (-0.22 \pm 0.39) \cdot \log\left(\frac{L_{\mathrm{X,\,char}}}{L_{\mathrm{X,\,char}}(M)}\right).
\end{equation}
Also, the negative slope suggests an anticorrelation, the large uncertainty of $\pm 0.39$ indicates that the sample size is too small in order to draw conclusions about the correlation of the residuals. The linear correlation coefficient of $r = -0.19 \pm 0.32$ points in this direction as well. This result is kind of expected since the observed anticorrelation in the larger sample of \citetalias{Flaischlen2021} is weak, as suggested by the models of X-ray driven photoevaporation. 

\begin{figure}
\resizebox{\hsize}{!}
        {
                \includegraphics[width=\hsize]{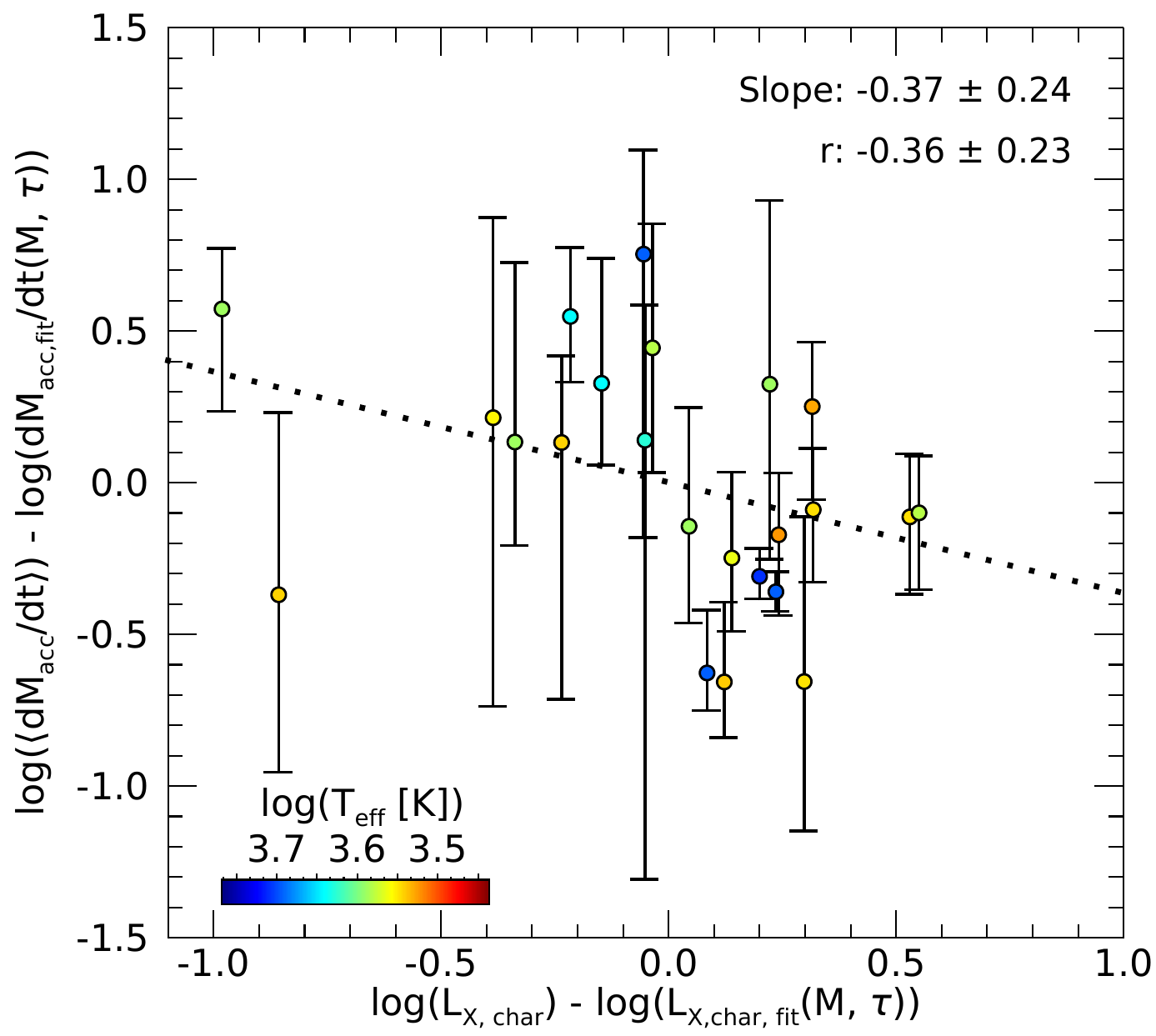}
        }
        \caption{Residual accretion rates vs. residual X-ray luminosities for the ONC taking both the mean mass and the mean isochronal age dependence into account. The dotted line shows the best fit obtained with LINMIX\_ERR. The solid bars indicate the maximum and the minimum residual accretion rates for each source.} \label{fig:figresfit}
\end{figure}

Since it is known that the accretion rates as well as the X-ray luminosities tend to decrease with increasing age \citep[e.g.,][]{Preibisch2005b, Telleschi2007, Manara2012}, it can be expected that the scatter in the relation can be reduced by correcting not only for the mass but also for the isochronal age. Therefore, we fitted planes to the quantities with a standard Levenberg-Marquardt regression and regarded both the mean mass and the mean isochronal age simultaneously. Uncertainties were not taken into account in this step. Then, we performed the partial regression with the residuals calculated from these relations and found:
\begin{multline}
\log\left(\frac{\langle\dot{M}_\mathrm{acc}\rangle}{\dot{M}_\mathrm{acc}(M, \tau) }\right) = \\ 0.00 \pm 0.09
+ (-0.37 \pm 0.24) \cdot \log\left(\frac{L_{\mathrm{X,\,char}}}{L_{\mathrm{X,\,char}}(M, \tau)}\right) \label{eq:eqfinal}
\end{multline}
with a linear correlation coefficient of $r = -0.36 \pm 0.23$. A scatter plot of the residuals together with the regression result is shown in Fig. \ref{fig:figresfit}.
Equation \ref{eq:eqfinal} shows a $1\sigma$ significant anticorrelation between the residual accretion rates and the residual X-ray luminosities, but the sample is too small to draw a stronger conclusion. The interpretation of this result is further difficult due to the uncertainties of the isochronal age estimates \citep[e.g.,][]{Preibisch2012, Soderblom2014} and their correlation with the accretion rates \citep{DaRio2014}.

\subsection{Influence of accretion variability on a larger sample}

In order to take advantage of a larger sample, we harkened back to our previous analysis in \citetalias{Flaischlen2021} and tried to incorporate our findings about the accretion variability in this study. To do so, we set the uncertainty of the accretion rates to twice the typical range of the accretion rate variability of $\sim 0.54~\mathrm{dex}$ we found in our monitoring survey and repeated the partial linear regression analysis. It is not likely that the sample accommodates targets with substantially larger variability. Young stars of FU Ori type are known to produce changes in the accretion rate for several orders of magnitude due to bursting events \citep{Hartmann2016}, but with typically $\sim 10^{-5}~M_\odot~\mathrm{yr}^{-1}$ \citep[e.g.,][]{Banzatti2015}, the accretion rates of these targets are more than one order of magnitude higher than the highest value measured in the HST sample used in \citetalias{Flaischlen2021}.

Nevertheless, we intended to include the possibility of such rare bursting events in our analysis and chose a Monte Carlo approach: conservatively, we assumed that $5~\%$ of the targets in our sample were subject to strong variability. Therefore, we assigned $16$ of the $322$ sources a scatter of $5~\mathrm{dex}$ and repeated the partial regression analysis. We iterated this process $1000$ times and randomly shuffled the $16$ sources to which the $5~\mathrm{dex}$ scatter has been assigned for every iteration step. Each time, we stored the linear correlation coefficients drawn from the \verb|LINMIX_ERR| posterior distributions, consisting of $12600$ values. From their distribution, we determined the mean and the standard deviation to obtain $r = -0.74 \pm 0.25$. We can conclude with $95~\%$ confidence that $-1.00 \leq r \leq -0.24$, meaning that the probability is $\sim 95~\%$ that the null hypothesis (no or a positive correlation) is ``not'' true. Thus, the anticorrelation is significant, even under the conservative assumption of strong accretion variability contamination of our sample.


\section{Summary and conclusions}
We presented the results of a multiyear photometric monitoring of accretion rates of young stars in the ONC. The observations were carried out with the WWFI instrument and comprises photometry in the $u^\prime$, $g^\prime$, and $r^\prime$ filters. Accretion rates were estimated from the observed displacement of the source positions in a color-color diagram by modeling the colors as a combination of an empirically determined, photospheric contribution and an accretion model.

The results allowed us to study how the accretion rate varies on timescales up to a few years. We found a typical interquartile range of $\sim 0.3~\mathrm{dex}$. We showed that the accretion rate variability has likely not introduced a bias in our previous study of the relation between X-ray activity and accretion rates \citep{Flaischlen2021}, where we reported a weak anticorrelation between the quantities, supporting the theoretical models of X-ray driven photoevaporation \citep{Ercolano2008a, Ercolano2009, Drake2009, Owen2010, Picogna2019}.

\begin{acknowledgements}
We wish to thank the referee for helpful suggestions. We thank the LMU master physics student Benedikt Mayr for his contribution to the WWFI data reduction.
This work made use of data obtained at the Wendelstein Observatory. The 2 m telescope project is funded by the Bavarian government and by the German Federal government through a common funding process. Part of the 2 m instrumentation including some of the upgrades for the infrastructure and the 40 cm telescope housing were funded by the Cluster of Excellence ``Origin of the Universe'' of the German Science foundation DFG.
This work was funded by the \emph{Deut\-sche For\-schungs\-ge\-mein\-schaft}
(DFG, German Research Foundation) under DFG project number 325594231 in the context of the
Research Unit FOR 2634/1: ``Planet Formation Witnesses and Probes:
TRANSITION DISKS''.
This project has received funding from the European Union's Horizon 2020 research and innovation programme under the Marie Sklodowska-Curie grant agreement No 823823 (DUSTBUSTERS).
This research was partly supported by the Excellence Cluster ORIGINS which is funded by the \emph{Deut\-sche For\-schungs\-ge\-mein\-schaft} (DFG, German Research Foundation) under Germany's Excellence Strategy - EXC-2094-390783311.
This work has made use of data from the European Space Agency (ESA) mission
{\it Gaia} (\url{https://www.cosmos.esa.int/gaia}), processed by the {\it Gaia}
Data Processing and Analysis Consortium (DPAC,
\url{https://www.cosmos.esa.int/web/gaia/dpac/consortium}). Funding for the DPAC
has been provided by national institutions, in particular the institutions
participating in the {\it Gaia} Multilateral Agreement.
This research has made use of the SIMBAD database
and the VizieR catalog services operated at Strasbourg astronomical Data Center
(CDS).
\end{acknowledgements}

%
%

\bibliographystyle{aa_url} 
\bibliography{AA202142630_final_sources} 

\begin{appendix} 
 \section{Observation log}
Table \ref{tab:tabepochs} lists how many targets were observed per epoch and filter.

 \begin{table}
\caption{\label{tab:tabepochs}Observed objects per epoch and filter.}
\centering
\begin{tabular}{crrr}
\hline\hline\\[-2.0ex]
Date (UT) & N$^{\circ}$ $u^\prime$\tablefootmark{(1)} & N$^{\circ}$ $g^\prime$\tablefootmark{(1)} & N$^{\circ}$ $r^\prime$\tablefootmark{(1)} \\
$[$YY-MM-DD$]$ & & & \\[0.4ex]
\hline\\[-1.6ex]
2014-11-24 & $ 136$ & $ 346$ & $ 553$ \\
2015-11-15 & $   0$ & $ 670$ & $ 821$ \\
2015-11-16 & $ 164$ & $   0$ & $   0$ \\
2015-12-11 & $   0$ & $   0$ & $ 841$ \\
2016-02-06 & $  51$ & $   0$ & $   0$ \\
2016-02-18 & $  71$ & $   0$ & $   0$ \\
2016-03-14 & $ 204$ & $   0$ & $   0$ \\
2016-03-16 & $   0$ & $ 438$ & $   0$ \\
2016-09-25 & $ 130$ & $ 375$ & $   0$ \\
2016-10-13 & $ 157$ & $ 425$ & $   0$ \\
2016-12-08 & $ 205$ & $ 598$ & $ 633$ \\
2016-12-11 & $   0$ & $ 652$ & $ 861$ \\
2016-12-16 & $ 239$ & $   0$ & $   0$ \\
2016-12-23 & $ 197$ & $ 521$ & $ 678$ \\
2016-12-31 & $ 216$ & $ 634$ & $ 782$ \\
2017-01-19 & $ 224$ & $ 405$ & $ 324$ \\
2017-01-22 & $ 231$ & $ 649$ & $ 826$ \\
2017-01-23 & $ 218$ & $ 617$ & $ 785$ \\
2017-01-24 & $ 219$ & $ 620$ & $ 739$ \\
2017-01-25 & $ 206$ & $ 616$ & $ 764$ \\
2017-09-28 & $ 225$ & $ 654$ & $ 847$ \\
2017-10-18 & $ 180$ & $   0$ & $   0$ \\
2017-12-25 & $ 121$ & $ 468$ & $ 602$ \\
2018-01-10 & $ 135$ & $   0$ & $   0$ \\
2018-01-26 & $ 200$ & $ 556$ & $ 733$ \\
2018-02-04 & $ 163$ & $   0$ & $ 765$ \\
2018-03-08 & $ 106$ & $   0$ & $   0$ \\
2018-09-30 & $ 234$ & $ 637$ & $   0$ \\
2018-10-05 & $ 221$ & $ 621$ & $ 780$ \\
2018-10-08 & $ 207$ & $ 625$ & $ 840$ \\
2018-10-09 & $   0$ & $   0$ & $ 705$ \\
2018-11-12 & $ 212$ & $ 631$ & $ 828$ \\
2018-11-13 & $ 210$ & $ 617$ & $ 582$ \\
2018-11-15 & $ 227$ & $ 657$ & $ 837$ \\
2018-11-16 & $ 216$ & $ 645$ & $ 839$ \\
2018-11-21 & $ 230$ & $ 655$ & $ 851$ \\
2018-11-22 & $ 186$ & $ 482$ & $ 462$ \\
2018-11-27 & $ 133$ & $ 470$ & $ 660$ \\
2018-12-02 & $ 124$ & $ 485$ & $ 689$ \\
2018-12-14 & $   0$ & $ 518$ & $ 790$ \\
2019-02-13 & $ 186$ & $ 601$ & $ 760$ \\
2019-02-17 & $   0$ & $ 586$ & $ 787$ \\
 \hline
\end{tabular}
\tablefoot{
\tablefoottext{1}{Amount of observed objects in the respective filter.}
}
\end{table}

\section{Derived filter properties}
For AB magnitude systems like the SDSS system WWFI uses, the zero-point $Z_\zeta$ is given by
\begin{equation}
Z_\zeta = 2.5 \log \left(\frac{\lambda^2_{p,\,\zeta}}{c}\right) + 48.6, \label{eq:eqZ}
\end{equation}
with the pivot wavelength $\lambda_{p,\,\zeta}$, which can be calculated via
\begin{equation}
\lambda^2_{p,\,\zeta} = \frac{\int \lambda \, T_{\zeta, \lambda} \, \mathrm{d} \lambda}{\int \lambda^{-1} \, T_{\zeta, \lambda} \, \mathrm{d} \lambda}, \label{eq:eqpivot}
\end{equation}
where $T_{\zeta, \lambda}$ is the instrumental response function of the respective filter and the system it is installed on \citep[]{Casagrande2014}. We used $T_{\zeta, \lambda}$ as given by \cite{Kosyra2014}.

We also calculated the ratio of the extinction through each filter and the visual extinction, $A_\zeta / A_V$. To this aim, we assumed a constant flux, an approximation justified in Appendix \ref{app:ext}, and the reddening law of \cite{Cardelli1989} as well as a galactic reddening parameter of $R_V = 3.1$, which was found to be appropriate for our sample \citep{DaRio2010}. Using Eq. \ref{eq:eqmag}, we calculated $\Delta m_{\zeta,\,A_V} := m_{\zeta,\,A_V} - m_{\zeta,\,A_V\,=\,0~\mathrm{mag}}$ for $A_V$ values between $0$ and $5~\mathrm{mag}$ in steps of $0.1~\mathrm{mag}$. Then, we fitted a straight line of origin through the $\Delta m_{\zeta,\,A_V}$ - $A_V$ relation and deduced $A_\zeta / A_V$ from the slope. The calculated parameters of the filter system are summarized in Table \ref{tab:tabfiler}.

\begin{table}
\caption{\label{tab:tabfiler}Filter properties.}
\centering
\begin{tabular}{cccr}
\hline\hline\\[-2.0ex]
Filter & $Z_P$\tablefootmark{(1)} $[\mathrm{mag}]$ & $\lambda_p$\tablefootmark{(2)} $[\mathrm{nm}]$ & \multicolumn{1}{c}{Extinction\tablefootmark{(3)}} \\[0.3ex]
\hline\\[-1.7ex]
$u^\prime$ & $20.1834$ & $352.44$ & $A_{u^\prime} / A_V = 1.5681$ \\
$g^\prime$ & $20.8146$ & $480.54$ & $A_{g^\prime} / A_V = 1.1625$ \\
$r^\prime$ & $21.3727$ & $621.64$ & $A_{r^\prime} / A_V = 0.8546$ \\
\hline
\end{tabular}
\tablefoot{
\tablefoottext{1}{Zero-point.}
\tablefoottext{2}{Pivot wavelength.}
\tablefoottext{3}{Calculated using the reddening law of \cite{Cardelli1989} and a galactic reddening parameter of $R_V = 3.1$.}
}
\end{table}

\section{Accuracy of the extinction estimates} \label{app:ext}

In the described approach, we assumed that extinction moves the points in the color-color diagram along lines parallel to the reddening vector. However, the reddening vectors are not exactly parallel. This would be the case if the flux density was constant over the filter's bandwidth, which is not the case in reality. In order to get more precise solutions, the term $10^{-0.4 A_\lambda}$ has to be multiplied to the total flux density defined by Eq. \ref{eq:eqflux}. We repeated the steps of section \ref{sec:method} for a set of $A_V$ values between $0$ and $10~\mathrm{mag}$, in steps of $0.01~\mathrm{mag}$. This renders basically Table \ref{tab:iso} into a cube, from which the photospheric magnitudes can be interpolated with reasonable precision for any $A_V$ in the chosen interval. Similarly, the colors associated with the accretion model can be obtained. 

For a given $T_\mathrm{eff}$, the missing two values $\eta$ and $A_V$ can be obtained by comparing the calculated colors with the observed ones and regarding the relations as a system of two nonlinear equations, which can be solved with a Quasi-Newton method, for instance. We found that the $A_V$ and $L_\mathrm{acc}$ values obtained with this more precise method differ not significantly from the values gained with the simpler method for $A_V < 5~\mathrm{mag}$. Since it is computationally much less expensive, we chose to use the simpler method for our analysis.

\section{The photospheric magnitudes} \label{app:iso}

\begin{figure}
\resizebox{\hsize}{!}
        {
                \includegraphics[width=\hsize]{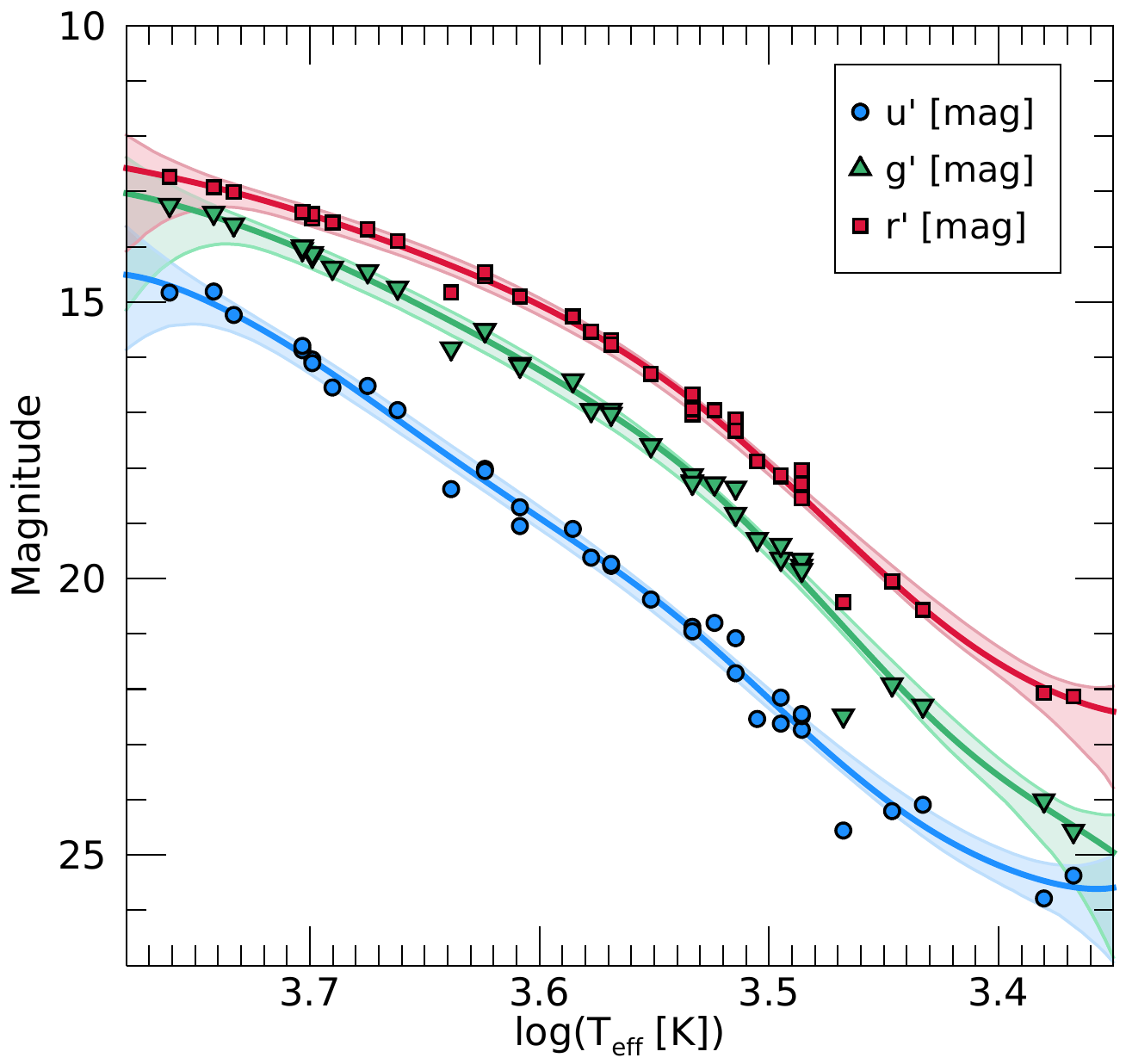}    
        }
        \caption{
				Results of the synthetic photometry with the photospheric template spectra observed by \cite{Manara2013,Manara2017}. Solid lines show the best fits and the shaded regions the $95~\%$ confidence interval. The lines can be recreated by cubic spline interpolation of the values listed in Table \ref{tab:iso}. Shown are the $u^\prime g^\prime r^\prime$ magnitudes as a function of $T_\mathrm{eff}$. 				
        } \label{fig:figphotteff}
\end{figure}

Figure \ref{fig:figphotteff} shows the magnitudes $m_{\mathrm{phot},\, u^\prime}$, $m_{\mathrm{phot},\, g^\prime}$, and $m_{\mathrm{phot},\, r^\prime}$ obtained from the photospheric template spectra as described in Sec. \ref{sec:methphototrack} as a function of $T_\mathrm{eff}$. The fitting results are listed in Table \ref{tab:iso}.

\begin{table}
\caption{\label{tab:iso}Magnitudes of non-accreting stars scaled to a radius of $R_\odot$ and a distance of $403~\mathrm{pc}$ as a function of $T_\mathrm{eff}$.}
\centering
\begin{tabular}{crrr}
\hline\hline\\[-2.3ex]
\multicolumn{1}{c}{$T_\mathrm{eff}$ [K]} & \multicolumn{1}{c}{$u^\prime$ [mag]} & 
\multicolumn{1}{c}{$g^\prime$ [mag]} & \multicolumn{1}{c}{$r^\prime$ [mag]} \\
\hline\\[-1.6ex]
$5835$ & $14.64_{-0.84}^{+0.53}$ & $13.16_{-1.20}^{+0.42}$ & $12.69_{-0.94}^{+0.39}$ \\[1.0ex]
$5669$ & $14.82_{-0.57}^{+0.34}$ & $13.30_{-0.74}^{+0.29}$ & $12.81_{-0.60}^{+0.26}$ \\[1.0ex]
$5505$ & $15.06_{-0.39}^{+0.22}$ & $13.47_{-0.46}^{+0.20}$ & $12.93_{-0.36}^{+0.18}$ \\[1.0ex]
$5339$ & $15.34_{-0.30}^{+0.16}$ & $13.66_{-0.31}^{+0.16}$ & $13.08_{-0.24}^{+0.14}$ \\[1.0ex]
$5174$ & $15.67_{-0.28}^{+0.14}$ & $13.87_{-0.26}^{+0.15}$ & $13.24_{-0.21}^{+0.13}$ \\[1.0ex]
$5010$ & $16.03_{-0.27}^{+0.15}$ & $14.11_{-0.26}^{+0.14}$ & $13.42_{-0.20}^{+0.13}$ \\[1.0ex]
$4845$ & $16.44_{-0.25}^{+0.17}$ & $14.38_{-0.25}^{+0.15}$ & $13.62_{-0.20}^{+0.14}$ \\[1.0ex]
$4680$ & $16.88_{-0.22}^{+0.19}$ & $14.68_{-0.25}^{+0.16}$ & $13.84_{-0.18}^{+0.15}$ \\[1.0ex]
$4514$ & $17.34_{-0.21}^{+0.21}$ & $15.00_{-0.24}^{+0.17}$ & $14.08_{-0.17}^{+0.16}$ \\[1.0ex]
$4350$ & $17.81_{-0.20}^{+0.22}$ & $15.35_{-0.23}^{+0.18}$ & $14.34_{-0.17}^{+0.16}$ \\[1.0ex]
$4184$ & $18.30_{-0.20}^{+0.22}$ & $15.72_{-0.22}^{+0.18}$ & $14.63_{-0.19}^{+0.15}$ \\[1.0ex]
$4019$ & $18.79_{-0.20}^{+0.21}$ & $16.13_{-0.22}^{+0.17}$ & $14.97_{-0.19}^{+0.13}$ \\[1.0ex]
$3855$ & $19.30_{-0.22}^{+0.18}$ & $16.58_{-0.23}^{+0.16}$ & $15.36_{-0.19}^{+0.12}$ \\[1.0ex]
$3689$ & $19.84_{-0.24}^{+0.14}$ & $17.08_{-0.25}^{+0.13}$ & $15.81_{-0.19}^{+0.10}$ \\[1.0ex]
$3525$ & $20.45_{-0.25}^{+0.11}$ & $17.66_{-0.25}^{+0.11}$ & $16.35_{-0.20}^{+0.09}$ \\[1.0ex]
$3360$ & $21.17_{-0.23}^{+0.11}$ & $18.36_{-0.25}^{+0.09}$ & $17.01_{-0.19}^{+0.08}$ \\[1.0ex]
$3195$ & $22.00_{-0.18}^{+0.15}$ & $19.22_{-0.23}^{+0.11}$ & $17.79_{-0.17}^{+0.10}$ \\[1.0ex]
$3029$ & $22.88_{-0.14}^{+0.23}$ & $20.24_{-0.18}^{+0.19}$ & $18.69_{-0.14}^{+0.16}$ \\[1.0ex]
$2865$ & $23.74_{-0.13}^{+0.30}$ & $21.34_{-0.16}^{+0.29}$ & $19.64_{-0.12}^{+0.24}$ \\[1.0ex]
$2700$ & $24.51_{-0.19}^{+0.34}$ & $22.45_{-0.22}^{+0.36}$ & $20.59_{-0.15}^{+0.31}$ \\[1.0ex]
\hline
\end{tabular}
\tablefoot{The values describe the best fit of a nonparametric regression and the $95~\%$ confidence interval. The input for the regression was synthetic $u^\prime g^\prime r^\prime$ photometry of X-Shooter spectra from diskless class III objects observed by \cite{Manara2013, Manara2017}.}
\end{table}

\section{Lines of constant accretion through total luminosity} \label{app:lines}

In Fig. \ref{fig:fig2cd2}, lines of constant ratio $L_\mathrm{acc} / L_\mathrm{tot}$ are shown. In order to 
calculate them, one can replace the factor $(R_* / R_\odot)^2$ in Eq. \ref{eq:eqlacc} with $L_* / L_\odot \cdot (T_\odot / T_\mathrm{eff})^4$ and
solve for $\eta$. Substituting $L_*$ in the resulting expression with $L_\mathrm{tot} - L_\mathrm{acc}$ yields
\begin{equation}
\eta\left(T_\mathrm{eff}\right) = \frac{1}{\left(L_\mathrm{tot} / L_\mathrm{acc} - 1\right) \cdot L_\mathrm{acc,\,0} / L_\odot} \cdot \left(\frac{T_\mathrm{eff}}{T_\odot}\right)^4.
\end{equation}

\end{appendix}

\end{document}